\documentclass{sig-alternate}

\usepackage{epsf}
\usepackage{epsfig}
\usepackage{amsmath, alltt, amssymb, xspace}
\usepackage{wrapfig}
\usepackage{subfigure}
\usepackage{cite}
\usepackage{subfigure}
\usepackage{graphicx}
\usepackage{mathrsfs}
\usepackage{algorithm}
\usepackage{algorithmic}

\usepackage{times}

\newtheorem{define}{Definition}[section]

\newcounter{task}
\begin{document}
%
\title{Deriving Latent Social Impulses to Determine Longevous Videos}

\author{
Qingbo Hu\\
       \affaddr{University of Illinois at Chicago, USA}\\
       \email{qhu5@uic.edu}
\and
Guan Wang\\
       \affaddr{University of Illinois at Chicago, USA}\\
       \email{gwang26@uic.edu}
\and
Philip S. Yu\\
      \affaddr{University of Illinois at Chicago, USA}\\
      \affaddr{King Abdulaziz University, Saudi Arabia}\\
       \email{psyu@uic.edu}       
}

\maketitle

\begin{abstract}

Online video websites receive huge amount of videos daily from users all around the world. How to provide valuable recommendations to viewers is an important task for both video websites and related third parties, such as search engines. Previous work conducted numerous analysis on the view counts of videos, which measure a video's value in terms of popularity. However, the long-lasting value of an online video, namely \emph{longevity}, is hidden behind the history that a video accumulates its "popularity" through time. Generally speaking, a longevous video tends to constantly draw society's attention. With focus on one of the leading video websites, Youtube, this paper proposes a scoring mechanism quantifying a video's longevity. Evaluating a video's longevity can not only improve a video recommender system, but also help us to discover videos having greater advertising value, as well as adjust a video website's strategy of storing videos to shorten its responding time. In order to accurately quantify longevity, we introduce the concept of latent \emph{social impulses} and how to use them measure a video's longevity. In order to derive latent social impulses, we view the video website as a digital signal filter and formulate the task as a convex minimization problem. The proposed longevity computation is based on the derived social impulses. Unfortunately, the required information to derive social impulses are not always public, which makes a third party unable to directly evaluate every video's longevity. To solve this problem, we formulate a semi-supervised learning task by using part of videos having known longevity scores to predict the unknown longevity scores. We propose a \emph{Gaussian Random Markov} model with \emph{Loopy Belief Propagation} to solve this problem. The conducted experiments on Youtube demonstrate that the proposed method can significantly improve the prediction results comparing to baseline models.

\end{abstract}

\section{Introduction}

Contemporary Internet world has witnessed the rise of numerous online video websites. According to the statistics, one of the leading online video websites, Youtube, is visited by over 800 million unique users in every month and receives 72 hours of uploaded videos in every minute\footnote{http://www.youtube.com/t/press\_statistics}. Such a huge amount of uploaded videos and visiting users have brought about both opportunities and challenges. Investors favor a video website with a large number of registered users and daily visits due to the potential advertising value behind it. For example, the most viewed video in the world, a Korean MV called Gangnam Style, is worth over 8 million U.S dollars in the online advertising deals\footnote{http://abcnews.go.com/blogs/technology/2013/01/psys-gangnam-style-is-an-8-million-blockbuster-hit-on-youtube/}. Therefore, on one hand, in order to attract viewers, video websites must excel in recommending valuable videos to its users. On the other hand, they may also solicit third party like search engines to help make video recommendations. 

With access to a registered user's profile, log and full information of uploaded videos, a video website is able to make recommendations that the user is potentially interested. However, such protected information is usually missing or incomplete to a third-party recommender system. For instance, once a search engine has a set of videos matching to the key words in a user's query, the limited information is usually insufficient to make highly valued recommendations. In this case, how to evaluate a video's value based on its public information becomes particularly important. A straightforward method to measure the value is based on a video's total views, i.e. {\bf \emph{popularity}}. However, it may be problematic to judge a video's value solely based on the times it has been watched, which will be introduced in the following paragraphs.

\subsection{Popularity v.s. Longevity}

\begin{figure}
	\centering
	\subfigure["Gabrielle Aplin - Please Don't Say You Love Me"] {
		\includegraphics[height=0.6in, width=85mm]{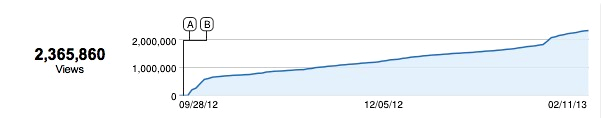}
		\label{fig: video_case_1}
	}	
	\subfigure[A repost of "Innocence of Muslims" ] {
		\includegraphics[height=0.6in, width=85mm]{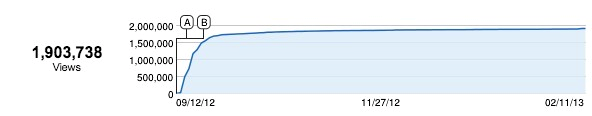}
		\label{fig: video_case_2}
	}
	
	\caption{Youtube Statistics of Two Videos (Feb. 2013)}
	\label{fig: video_cases}
\end{figure}

Previously, researchers have proposed numerous approaches to evaluate the popularity of online videos by directly using their views. The concept of popularity usually depends on a video's views at a specific time point~\cite{cha2007tube, Youtube_statistics, Youtube_popularity_growth, Szabo_2010, Pinto2013}. However, another perspective of a video's value, which is the long-lasting value, a.k.a. \emph{longevity}, has been hidden behind the view count growth pattern. In recent years, video sharing websites such as Youtube offer detailed information of how a video accumulates its views through time (such as Figure \ref{fig: video_cases}). 
Such statistical data provides us an opportunity to measure and evaluate a video's longevity. Roughly speaking, longevous videos have view count growth patterns that demonstrate promising long-term value, i.e. the ability to continuously attract society's attention. Therefore, a longevous video tends to have much higher view count increase in the future comparing to non-longevous ones. Accordingly, evaluating the longevity of videos have three major benefits: (1) recommending longevous videos to users can improve a website's recommender system; (2) longevous videos have higher advertising value since they are constantly watched; (3) evaluating the longevity can help adjust a video website's data storage strategy to shorten its responding time: longevous videos should be stored in quickly accessible locations, since they are frequently visited. Completely ignoring the longevity and only using the popularity to rank the value of online videos may cause problems. Next, we use two real online videos on the Youtube to show that. 

\begin{figure}
	\centering
	\subfigure["Gabrielle Aplin - Please Don't Say You Love Me"]{
		\includegraphics[height=10mm, width=75mm]{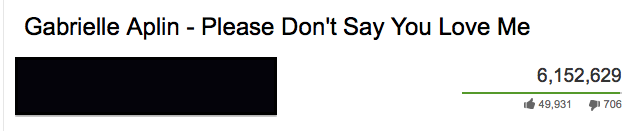}
	}
	\subfigure[A repost of "Innocence of Muslims"]{
		\includegraphics[height=10mm,width=75mm]{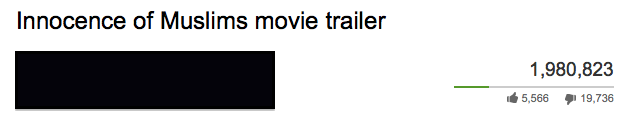}
	}
	\caption{Views of Two Videos (Sep. 2013)}
	\label{fig: video_cases_future}
\end{figure}

Figure \ref{fig: video_cases}\footnote{The figures are extracted in Feburary, 2012, which are the old version figures of Youtube statistical data. The marks of "A" and "B" in the Youtube statistics are the occurrence of related events, such as the time when it is firstly referred in a social network. They are irrelevant to our problems.} displays the Youtube statistical data of two videos at the end of February, 2013. Both of videos are published in the September, 2012 and accumulated around 2 million views. The first video is a song from a Pop singer, Gabrielle Aplin, and the second video is a repost of a short movie trailer insulting Muslins, which is speculated to trigger the incident of the assassination of U.S ambassador in Libya on September 11, 2012. By solely comparing the popularity (a.k.a. views), one may have to accept the fact that these two are roughly similar. However, the difference between them is actually easy to be discovered for human by just comparing the shape of curves in Figure \ref{fig: video_cases}: view count in Figure \ref{fig: video_case_1} continuously increases in the entire history, while Figure \ref{fig: video_case_2} only increases at the very early stage). In fact, due to such difference, the future views of the videos are totally different: in the September, 2013, the first video's views reach more than 6 million, while the second one is still less than 2 million (see screenshots in Figure \ref{fig: video_cases_future}). Therefore, without considering longevity, we may accidentally recommend a popular yet valueless (even malicious) movie to viewers in the Febuary, 2013. Are we able to develop a scoring system to rank different view count growth patterns? This paper exactly fulfills this quest. The proposed longevity calculation outputs a much higher score for the first video than the second one. In other words, the method judges the first video has a greater long-lasting value than the second one.

\subsection{Latent Temporal Bias and Social Impulses}

\begin{figure}[h]
	\centering
	\includegraphics[height=40mm, width=\linewidth]{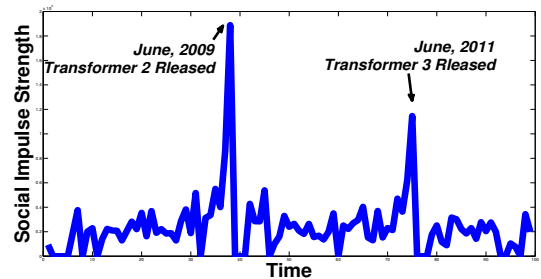}
	\caption{Derived Social Impulses for the First "Transformers" Movie's Trailer}
	\label{fig: transformer}
\end{figure}

A straightforward definition of a video's longevity can be directly derived from a video's {\bf \emph{view count's history}}. To eliminate ambiguity, we refer "view count's history" to the curve in which every point's value is the view count's difference between two adjacent time points in a video's statistical data (curves in Figure \ref{fig: video_cases} are examples of a video's statistical data). As a result, a naive definition of a video's longevity is the number of points having non-zero values in the view count's history, i.e. how many time intervals in the video's statistical data have non-zero view count increase.
However, we argue that the definition based on view count's history will be inaccurate because of the existence of {\bf \emph{latent temporal bias}}. Latent temporal bias is caused by the phenomenon that once a trend in the social media becomes unpopular, its influence will gradually decay instead of suddenly vanish~\cite{asur2011trends}. One of the possible reasons to cause such phenomena is the delay of user's response to an event in social networks~\cite{Saito2010}. Therefore, even though we may observe continuous view count increase in history, there may be only a few number of trends stimulating such increase. In this paper, we name these latent stimulus "trends" as {\bf \emph{social impulses}} and will provide a formal definition in Section 3.
Although social impulses are latent factors and reasons causing them are not always observable, some major ones can be explained by related events in real life. For instance, Figure \ref{fig: transformer} shows the social impulses derived by our method for the first "Transformers" movie's trailer on the Youtube. As we can see, the two major social impulses actually correspond to related real events: release of sequel movies.

\begin{figure}[h]
	\centering
	\subfigure[Google Search Trend of "Simon's Cat"] {
		\includegraphics[height=30mm, width=\linewidth]{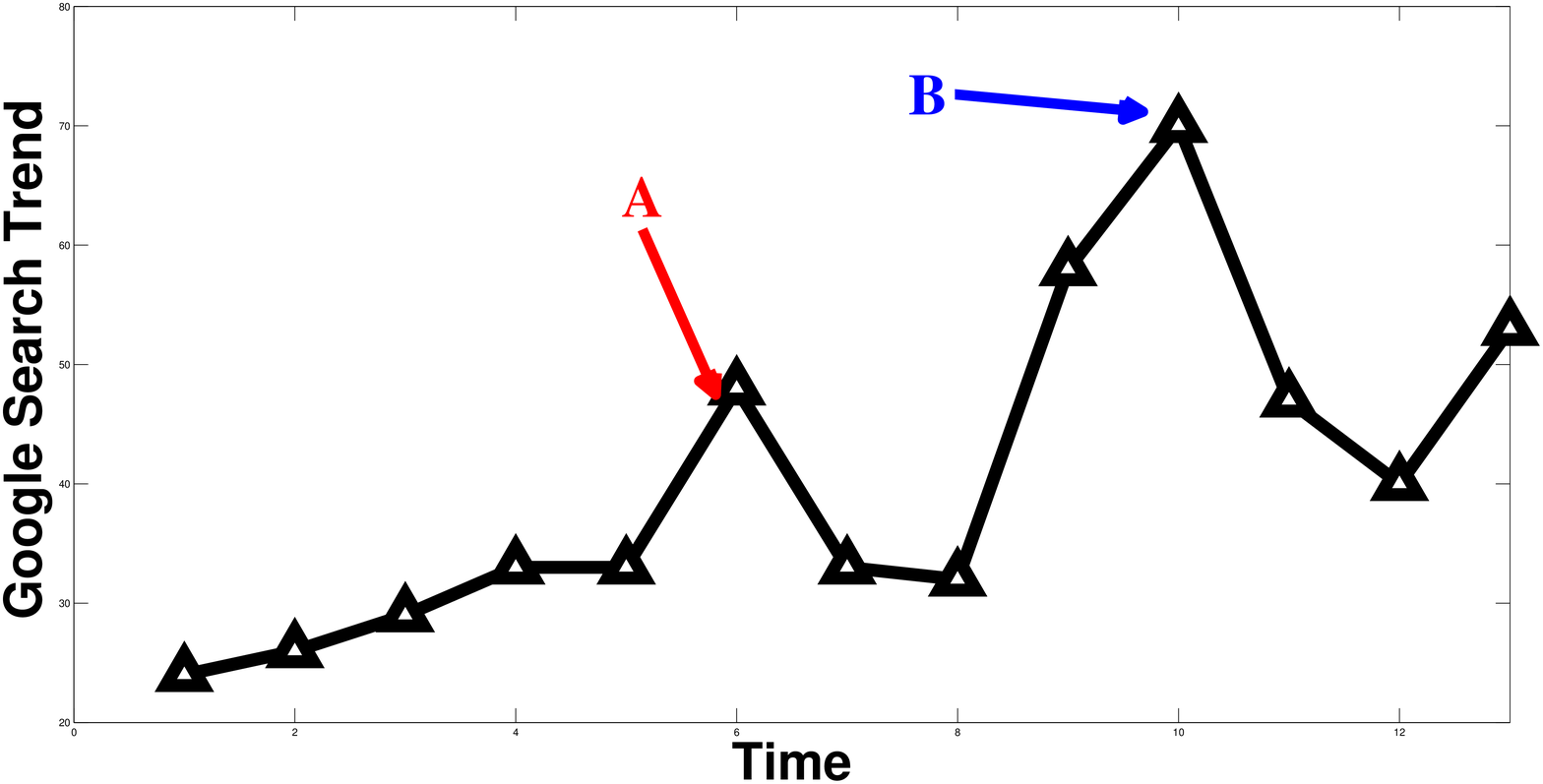}
		\label{fig: video_case_3}
	}
	
	\subfigure[Detected Social Impulses v.s. View Count Increase of "Cat Chat - Simon's Cat"] {
		\includegraphics[height=25mm, width=\linewidth]{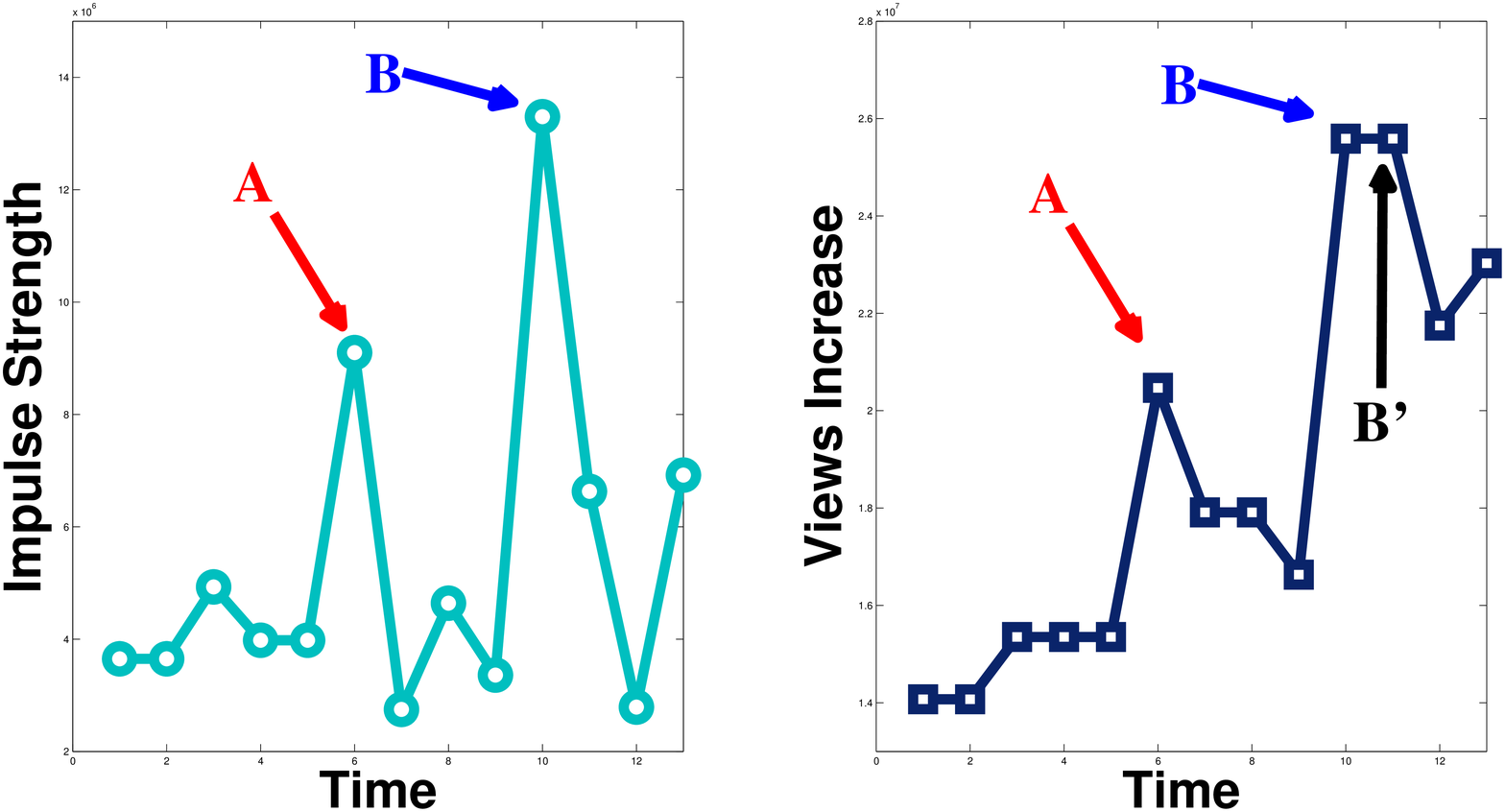}
		\label{fig: video_case_4}
	}
	
	\caption{Latent Temporal Bias in the view count's history of "Simon's Cat"}
	\label{fig: video_cases_2}
\end{figure}

We argue that using the history of social impulses to define longevity will be more reasonable, since it can more or less correct the latent temporal bias in the view count's history. Next, we will use a real-world case to demonstrate the importance of correcting such bias. Figure \ref{fig: video_case_3} is the Google Search trend for the keyword "Simon's Cat", which is an animated cartoon series by Simon Tofield. Since events related to "Simon's Cats" are not reported, we use the Google Search trend to infer the existence of such events. Figure \ref{fig: video_case_4} are the derived social impulses and the view count's history of an episode of "Simon's Cat", respectively. These three figures are all scaled to the same time interval and each time point corresponds to the data in a 7-day time window. As one can see, Google search trend has two obvious peaks at point A and B in Figure \ref{fig: video_case_3}. On the one hand, due to the latent temporal bias, view count increase decays slowly after one peak arrives and even causes the view at point B' to be the same as point B. It is more reasonable to consider a large amount of views at B' are only "side effects" of the event at point B. On the other hand, the history of derived social impulses better captures the major events implied by Google Search trend. 

How to discover social impulses becomes one of the major challenges we need to deal with. In the proposed model, we view the Youtube website as a digital signal filter, where the social impulses are input signals and the view count's history is the output. Moreover, we define the system's \emph{impulse response function} has an exponential form, which means that an input signal imposed on the system will have an exponentially decay effect on the outputs. Such definition is based on computational models describing user's behavior when respond to an event~\cite{Saito2010, Iwata2013}. In this way, we are able to formulate the task of discovering social impulses as a convex optimization problem and solve it. The proposed longevity scoring mechanism is based on the derived social impulses. 

\subsection{Prediction of Longevity.}

As stated previously, the longevity can be quantified based on the history of social impulses, which is derived by correcting the latent temporal bias in the view count's history. However, in our crawled dataset, we find that approximately 35\% of the videos do not have public view count's histories. The major reason for this is that many posters have turned off the function of displaying videos' statistical data. As a result, even though measuring a video's value is more important to the success of a third-party recommender system (since it usually does not have the viewer's information), a third party is still not able to directly evaluate the longevity for those videos lacking of view count's history. In fact, this is the second major problem we need to solve. In this paper, we investigate the possibility of using public information to predict a video's longevity, when its view count's history is missing. 

As a good news, we discover that most features of a video, such as total views, total shares and etc., are usually retrievable. Moreover, the list of a video's related videos (a.k.a. videos having related content) is also available. Therefore, we are able to form a connected network containing each video as a node. The nodes in the network are linked through edges representing "related videos" relationship. Moreover, since the longevity score of some nodes are known (their view count's histories are observable), these nodes can serve as labeled data. Thus, we formulate the problem of predicting a video's longevity score as a semi-supervised learning task in a network. Based on the problem formulation, we bring about two baseline methods that one can easily come up with. Furthermore, we also propose a {\bf \emph{Gaussian Markov Random Field (GMRF)}} model with {\bf \emph{Loopy Belief Propagation (LBP)}} to solve this semi-supervised problem. We conduct extensive experiments on a general video network, which is mixed with videos from different categories, as well as categorized video networks, which only contain videos in the same category. The results show that the proposed GMRF model with LBP can significantly outperform the baselines in terms of the Mean Squared Error (MSE).

We sum up our main contributions as follows:

\begin{itemize}
	\item To the best of our knowledge, we are the first ones to point out the importance and benefits of evaluating a video's long-lasting value, namely longevity. 
	\item Because of the latent temporal bias, instead of directly quantifying longevity based on the view count's history, we argue that it is better to be based on the history of latent social impulses. By viewing an online video website as a digital signal filter, we formulate the problem of deriving latent social impulses as a convex minimization problem and solve it. 
	\item We propose a simple yet useful longevity scoring mechanism based on the derived social impulses. Furthermore, we conduct statistical analysis and case study on a real-world dataset to support the rational and analyze the property of the proposed method to compute longevity scores.
	\item To address the problem of assessing a video's longevity when its view count's history is not observable to third parties, we formulate a semi-supervised task to predict longevity. We propose a solution by using a Gaussian Markov Random field model with Loopy Belief Propagation. The experimental evaluation on the real-world dataset confirms the proposed model is significantly better than compared baselines.
	
\end{itemize}

The remaining paper is organized as follows. In Section 2, we introduce the techniques we used to collect data from Youtube and basic statistics of the crawled dataset. In Section 3, we propose some basic definitions and the approach to derive social impulses. Then, we introduce the method to measure a video's longevity. In Section 4, we formulate the semi-supervised learning task of predicting a video's longevity score and introduce the proposed Gaussian Markov Random Field model with Loopy Belief Propagation. Experimental evaluation of the proposed model is also in this section. In Section 5, related work will be introduced. Finally, Section 6 will conclude this paper.

\section{Dataset Description}

In this section, we introduce how we collect our dataset from Youtube and basic statistics of the crawled dataset. In order to derive social impulses to quantify the longevity of a video and further formulate the prediction tasks, we need two types of data. The first one is a video's public features that displayed in the Youtube website, such as accumulated view count, favorite count, etc. The second type is the view count's history. This is used to calculate the longevity score of the video. The final obtained dataset contains a subset of popular (in terms of accumulated view count) videos that are published in 2010.

\subsection{Collecting Public Features} 

We use Youtube API 2.0\footnote{https://developers.google.com/youtube/} to directly collect this part of data. We design the crawling process to target videos categorically\footnote{The category is an inherent property of videos stored by Youtube.}. For each category, starting from some most viewed videos that are published in 2010, we initiate a BFS through the "related videos" relationship to collect more videos published in the same category and same year. The search stops until the size of the category reaches a predefined threshold (100,000 in our experiment). After that, we remove unpopular videos, whose accumulate view counts are less than a threshold (30,000 in our case), as well as those videos which do not have public view count's history. We run this process repeatedly by changing the targeting category. As a result, we obtain videos in six different categories: {\bf \emph{Animals, Music, News, Nonprofit, Sports}} and {\bf \emph{Technology}}, with total number of 65,016 videos and 117,370 edges\footnote{The final crawled edges are updated to connect videos not only within the same category, but also across different categories.} (related video relationship) connecting them. Table \ref{tab: crawled_data_1} elaborates all the public features we have gathered in this step, and Table \ref{tab: crawled_data_2} lists some basic statistics of the crawled dataset. Since sports and music stars often attract more viewers, it is not surprising to see that the dataset is biased towards "Sports" and "Music" videos, while "Animals" and "Nonprofit" videos are much less. The huge difference (more than ten times) between the average value of "likes" count and "dislikes" count indicates that the viewers tends to give more positive opinions than negative ones. This is probably caused by that a viewer does not want to offend (mark "dislikes") a video poster unless him/herself is seriously offended. The average length of a video in our dataset is less than 5 minutes, which is consistent with the statistics in~\cite{cha2007tube, Youtube_statistics}. 

\begin{table} [h]
	\centering
	\begin{tabular} {c | c}
		{\bf Feature} & {\bf Meaning}\\
		\hline
		id & Youtube ID of the video\\
		\hline
		category & category label of the video\\
		\hline
		viewCount & accumulated views of the video\\
		\hline
		favoriteCount & \# of users added the video in the favorite lists\\
		\hline
		averageRating & the average rating (1 to 5) received from viewers\\
		\hline	
		length & the length of the video (in seconds)\\
		\hline
		"like" count & \# of "Like"s clicked by viewers\\
		\hline
		"dislike" count & \# of "Dislike"s clicked by viewers\\
	\end{tabular}
	
	\caption{Collected Videos' Public Features}
	\label{tab: crawled_data_1}
\end{table}

\begin{table}[h]
	\centering
	\scriptsize
	\begin{tabular}{c | c | c | c | c | c | c }
		\hline
		{\bf Total Videos} & \multicolumn{2}{c |}{\bf Total Video Edges} & \multicolumn{2}{c |}{\bf Connected Components} & \multicolumn{2}{c }{\bf Network Diameter}\\
		\hline
		65,016 &  \multicolumn{2}{c |}{117,370} & \multicolumn{2}{c |}{2,365} & \multicolumn{2}{c }{27}\\
		\hline
		\hline
		{\bf Category} & Animals & Music & News & Nonprofit & Sports & Tech.\\
		\hline
		\# videos & 5,918 & 15,558 & 9,674 & 2,394 & 20,018 & 11,454\\
		\hline
		\hline
		& \multicolumn{2}{c |}{\bf viewCount} &  \multicolumn{2}{c |}{\bf favoriteCount} &  \multicolumn{2}{c}{\bf averageRating}\\
		\hline
		Avg. value & \multicolumn{2}{c |} {5000620.81} &  \multicolumn{2}{c |}{978.8} &  \multicolumn{2}{c}{4.404}\\
		\hline
		& \multicolumn{2}{c |}{\bf length (seconds)} &  \multicolumn{2}{c |}{\bf "like" count} &  \multicolumn{2}{c}{\bf "dislike" count}\\
		\hline
		Avg. value& \multicolumn{2}{c |}{266.95} &  \multicolumn{2}{c |}{1162.51} &  \multicolumn{2}{c}{95.18}\\
		\hline		
	\end{tabular}
	
	\caption{Statistics of the Youtube Dataset}
	\label{tab: crawled_data_2}
\end{table}

\subsection{Collecting View Count's Histories}

In order to collect the view count's history of a video, we apply the same method introduced in~\cite{Youtube_popularity_growth}. When Youtube generates the figure displaying the view count's history of a video, it calls a URL provided by Google charts API\footnote{The url for generating the figure of statistical data has changed in the current Youtube website. When we crawl the dataset, we are still able to using the same url address described in \cite{Youtube_popularity_growth}}. We analyze the result from the same URL for each video and extract all 100 points\footnote{The current Youtube statistical data contains more points} from them. These points are represented as the percentage of the total views. Therefore, by multiplying the final accumulated views to these percentage numbers, we are able to observe how view count is accumulated in history (such as curves in Figure \ref{fig: video_cases}). By taking the difference between the views of two adjacent time points, we obtain the view count's history, which contains 99 points denoting the view count increments.

\section{Social Impulses and Longevous Videos}

As we stated previously, once a social impulse occurs, it may cause continuous effect on the view count's history. Therefore, we define the longevity of a video based on the history of social impulses, instead of directly using view count's history. In this section, we introduce some basic definitions, our approach to derive social impulses and the method to compute longevity. After that, we use statistical analysis and a case study to demonstrate how the longevity can offer information of a video's long-lasting value. 

\subsection{Social Impulses}

\subsubsection{Derivation of Latent Social Impulses}

To be more clear, we first define some basic concepts. It is natural to view both social impulses and view count's history are series of non-negative digital signal at discrete time steps.
Therefore, we can further treat a video website as a time-invariant digital signal filter to model the relationship between social impulses and view count's history.
Formally, this filter is named as {\bf \emph{video signal filter}}, which has a mathematical definition according to the basic knowledge in the digital signal processing area:

\begin{define} [Video Signal Filter]
	For a given non-negative vector {\bf x} = $(x_0,x_1, ... x_N)$, a video signal filter is a function of mapping {\bf x} to another non-negative vector, {\bf y} = $(y_0, y_1, ... y_N)$, with the following discrete convolution holds: 
	\begin{equation*}
		y_{k (k =\{0,1,...,N\})} = \sum_{i=0}^{k}{x_i \cdot h_{k-i}}
	\end{equation*}
	$h_{i'}$ is the filter's Impulse Response Function (IRF), which maps $i' (i'={0,1,...n})$ to a non-negative real number.  
	
	\label{def: vsf}
\end{define}

Accordingly, social impulses and view count's history have the following mathematical definition, respectively:

\begin{define}[Social Impulses]
	Social impulses are input signals to a video signal filter, which is a non-negative vector. To be consistent with Def. \ref{def: vsf}, we denote them as {\bf x} = $(x_0,x_1, ... x_N)$.
\end{define}

\begin{define}[View Count's History]
	A view count's history is a series of output signals from a video signal filter, which is a non-negative vector. To be consistent with Def. \ref{def: vsf}, we denote it as {\bf y} = $(y_0, y_1, ... y_N)$.
	
	\label{def: vch}
\end{define}

In order to derive social impulses (inputs) according to view count's history (outputs), we first need to define the IRF in Def. \ref{def: vsf}. We use a exponential function with a tunable parameter as the IRF, since it is reasonable to assume that the effect of a single impulse to the system will exponentially decay as time goes by. Formally, for $i \in [0,N]$
, we assume $h_i$ has the following form:
	\begin{equation*}
		h_i = e^{- \gamma \cdot i}
	\end{equation*}
where $\gamma$ is a tunable parameter that defines how quickly the effect of a single social impulse decays. With a smaller value of $\gamma$, the effect decays more slowly. As a result, the derived social impulses are the more significant ones, which means the input signals tend to be sparser. Otherwise, the derived social impulses will be denser. 

As a result, we have a series of observed output signals (actual view count's history from dataset) and the predefined IRF, while the input signals (social impulses) are need to be derived. Therefore, if we denote {\bf $y^*$} = $(y_0^*,y_1^*,...y_n^*)$ as the observed view count's history, deriving social impulses can be formulated as a constrained Least Square Estimation problem:
	\begin{equation*}
	\begin{split}
		& min_{x_i}: \sum_{i=0}^{N} [y_i^* - y_i]^2 \\
		& subject\ to: \ \forall i, x_i \geq 0
	\end{split}
	\end{equation*}
For simplicity, we rewrite the problem in the following matrix form:
	\begin{equation*}
	\begin{split}
		& min_{\bf x}: \| {\bf y^*} - {\bf H} {\bf x} \|_2^2 \\
		& subject\ to: \ {\bf x} \succeq 0
	\end{split}
	\end{equation*}
{\bf H} is the matrix representing IRF, which is a lower triangular matrix:
	\begin{equation*}
		h_{i,j}=\left \{
			\begin{array}{lr}
				e^{- \gamma \cdot (i-j)} & i \geq j\\
				0 & i<y
			\end{array}
		\right.
	\end{equation*}
As we can see, since $\gamma$ is a constant we predefined, {\bf H} will be a matrix containing non-negative constants, and the above problem becomes a standard convex minimization problem. Therefore, we can employ the method in~\cite{lawson1974solving} to efficiently obtain a global optima {\bf x} for any given output signal ${\bf y^*}$. Table \ref{tab: symbols} lists all the important notations we use in this paper.

\begin{table}
	\scriptsize
	\centering
	\begin{tabular} {| c | c |}
		\hline
		{\bf Notation} & {\bf Meaning}\\
		\hline
		{\bf x} & Input signals (social impulses) to a video signal filter\\
		\hline
		{\bf y} & Output signals (view count's history) from a video signal filter\\
		\hline
		{\bf y*} & Observed output signals (view count's history) from the  dataset\\
		\hline
		$h_i$ & Impulse response function at step $i$\\
		\hline
		{\bf H} & Matrix representation of impulse response function\\
		\hline
		$\gamma$ & A tunable parameter in the impulse response function\\
		\hline
		$L(*)$ & Longevity Functions. $L: {\bf x} \rightarrow \mathbb{R}$\\
		\hline
		${\bf v_i}$ & The feature vector representation of video $i$\\
		\hline
		$v_{ik}$ & $k^{th}$ feature's value of video $i$\\
		\hline
		$w_{i,j}$ & Weight of the edge connecting $i$ and $j$\\
		\hline
		$s_i$ & Longevity score of video $i$\\
		\hline
		$\psi_i(*)$ & Node potential function of $i$\\
		\hline
		$\psi_{ij}(*)$ & Edge potential function of the edge connecting $i$ and $j$\\
		\hline
		$m_{i \rightarrow j}$ & Message sent from $i$ to $j$ in the Loopy Belief Propagation\\
		\hline
		$bel_i$ & Beliefs of the node $i$ in the Loopy Belief Propagation\\
		\hline
		
	\end{tabular}
	\caption{Important Mathematical Notations}
	\label{tab: symbols}
\end{table}

\subsubsection{Impulse Response Functions with Different $\gamma$}

\begin{figure}[h]
	\centering
	\includegraphics[width=\linewidth,height=35mm]{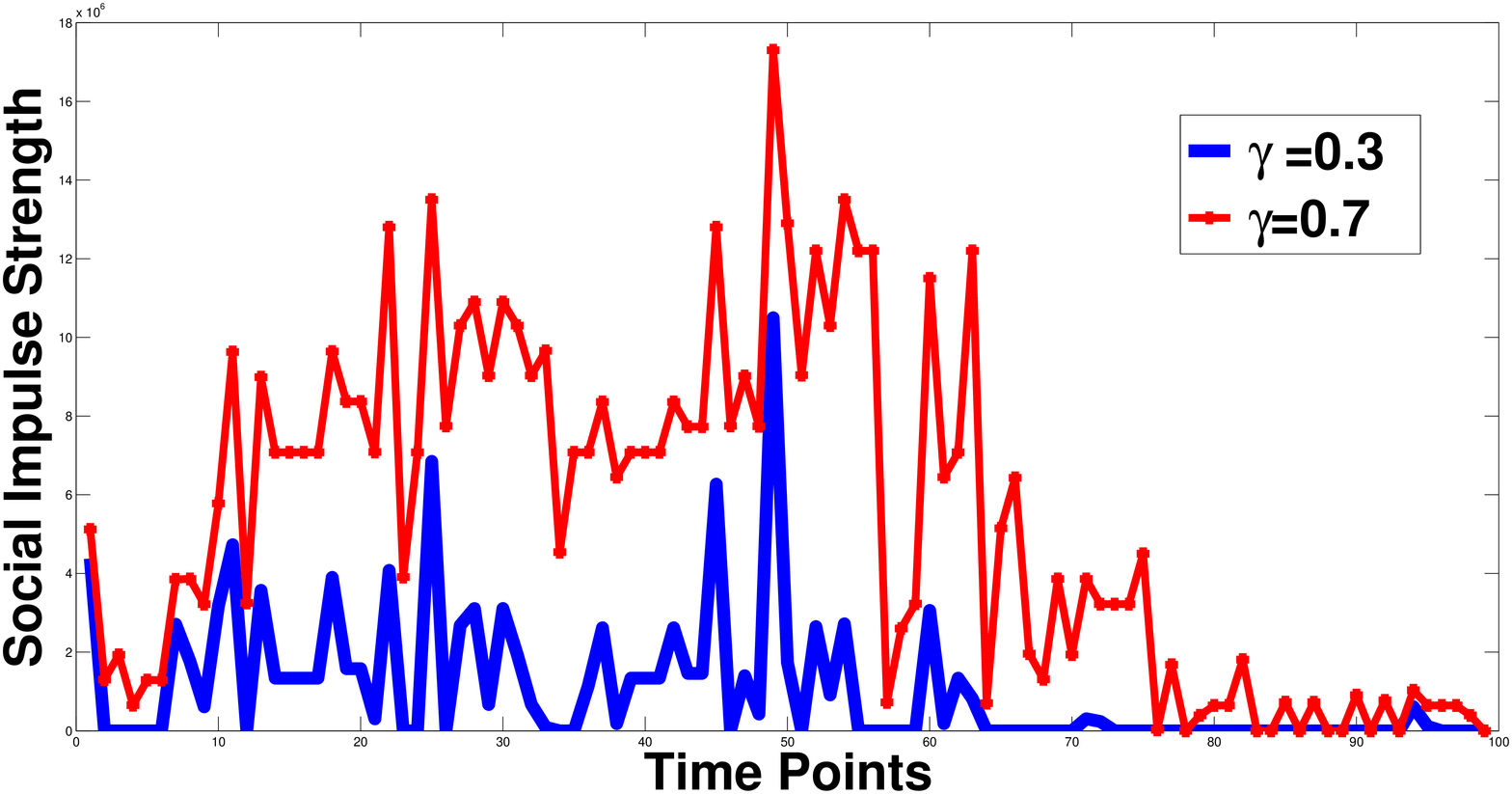}
	\caption{Effect of different $\gamma$}
	\label{fig:gamma_change}
\end{figure}

In this subsection, we analyze how the change of $\gamma$ value affects the derived social impulses. By changing the value of $\gamma$, we can obtain different IRFs. Such adjustment directly relates to the significance of the derived social impulses. As we stated previously, with a small $\gamma$ value, the obtained input signals vector, {\bf x}, tends to be sparser. We will use the same example introduced in Figure  \ref{fig: video_cases_2} to show the influence of the change of $\gamma$ value. Figure  \ref{fig:gamma_change} displays the derived social impulses of the video in Figure  \ref{fig: video_cases_2} with different $\gamma$ values. It is straightforward to see that with a smaller $\gamma$, derived social impulses are more isolated (sparser solution of {\bf x}). Moreover, the strength of them are weaker compare to a larger $\gamma$ value. This is easy to understand: an early impulse have more significant influence to the future under a smaller $\gamma$, so the effect of later derived social impulses are considered weaker. Furthermore, it is not hard to imagine that the change of $\gamma$ will also affect the computed longevity scores. which will be introduced later.

\subsection{Discovering Longevous Videos}

\subsubsection{Computation of Longevity}

Intuitively, longevous videos refer to those videos which can continuously attract society's interest. In other words, a longevous video tends to receive more social impulses than others. In order to quantify such abstract concept, we introduce a family of functions, namely \emph{{\bf Longevity Functions}}:

\begin{define}[Longevity Functions]
	Longevity functions is a family of functions that map social impulses to a real number. Formally, a longevity function is denoted as $L({\bf x}): {\bf x} \rightarrow \mathbb{R}$.
\end{define}

The values returned by a longevity function are called \emph{{\bf Longevity Scores}}, which can be used to compare the longevity of different videos. In this paper, we construct the longevity function by considering two important factors related to the intuition of longevity. (i) Longevity refers to the videos which have more active social impulses. Therefore, instead of incorporating the strength of social impulses in the longevity function, we should use the occurrence of social impulses. 
(ii) Social impulses occur later should be more valuable, since it implies that even after a long time when a video is published, it can still attract viewer's attention. As a result, we first propose $r_i$ to evaluate the longevity based on $x_i$:
\begin{equation}
	r_i = \sum_{i=0}^{N} [\mathbb{I}_{\{z | z \geq \epsilon \}}(x_i) \cdot (1+log(1+i))]
	\label{eq: ri}
\end{equation}
where $\mathbb{I}_{\{z | z \geq \epsilon \}}(x_i)$ is an indicator function indicating the occurrence of a social impulse at the time $i$, and $\epsilon$ is used to filter out possible noise. $(1+log(1+i))$ is a weighting term providing larger values to later social impulses. We use a log function to avoid the weight dominating $r_i$. Then, in order to simplify the computed longevity scores, we use a integer scale from 0 to 100 based on $r_i$: we compute the percentage number of each $r_i$ to the largest one and further round the result to the closest integer. Therefore, the proposed longevity function is written as follows:

\begin{equation}
	L_1({\bf x}) = Round(\frac{r_i} {max_i \{r_i\}} * 100)
	\label{eq: L1}
\end{equation} 
 
At last, the proposed longevity function, $L_1$, by Equation (\ref{eq: L1}) outputs a score from 0 (least long-lasting value) to 100 (greatest long-lasting value) for each video. In practice, if $\gamma$ is allowed to be changed, we can set $\epsilon$ = 0. The reason is that by tuning the value of $\gamma$ we can also achieve the goal of filtering out insignificant social impulses (noise). Therefore, in the sequel of this article, we set $\epsilon = 0$ in the experiments and analysis unless we specify otherwise. 

\subsection{Support and Analysis of $L_1$ Function}

\begin{figure}
	\centering
	\includegraphics[width=\linewidth]{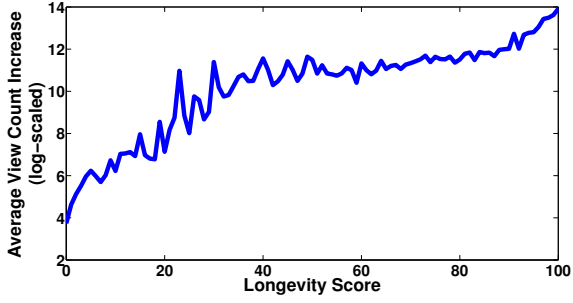}
	\caption{(Log-scaled) Average View Count Increase of Videos with Same Longevity Score ($\gamma$ = 0.3)}
	\label{fig:avg_view_score}
\end{figure}

To support that the proposed longevity function can well capture a video's long lasting value, we use the similar method in the examples demonstrated in Figure \ref{fig: video_cases_future}. First of all, we retrieve the view counts of all videos in the September, 2013, which is more than 10 months later since we crawled the original dataset. Afterwards, for videos having the same longevity score, we compute the average view count increase in this time gap. Therefore, such results reflect how a video's longevity score computed by $L_1$ correlates to its future view count increase. Figure \ref{fig:avg_view_score} demonstrates the results we obtained when $\gamma$=0.3, and the results for other $\gamma$ values are similar. Generally speaking, $L_1$ favors videos that can accumulate more views in the future, which reflects a video's long lasting value. Such observation confirms the rational of $L_1$.

\begin{figure} [h]

	\subfigure[Accumulated Views of a non-Longevous video] {
		\includegraphics[width=40mm, height=20mm]{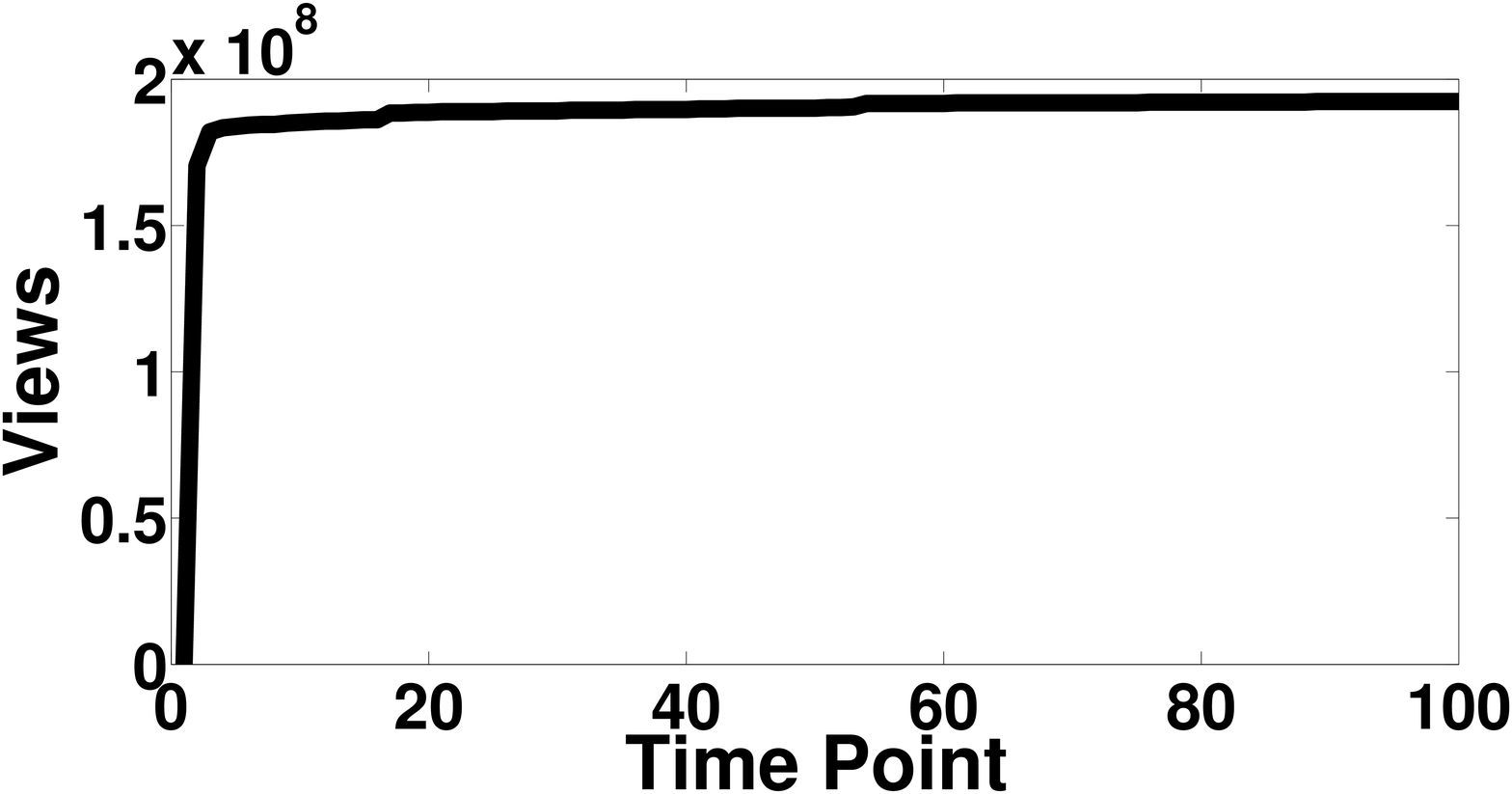}
		\label{fig: non_longevous}
	}
	\subfigure[Social Impulse History of a non-Longevous video] {
		\includegraphics[width=40mm, height=20mm]{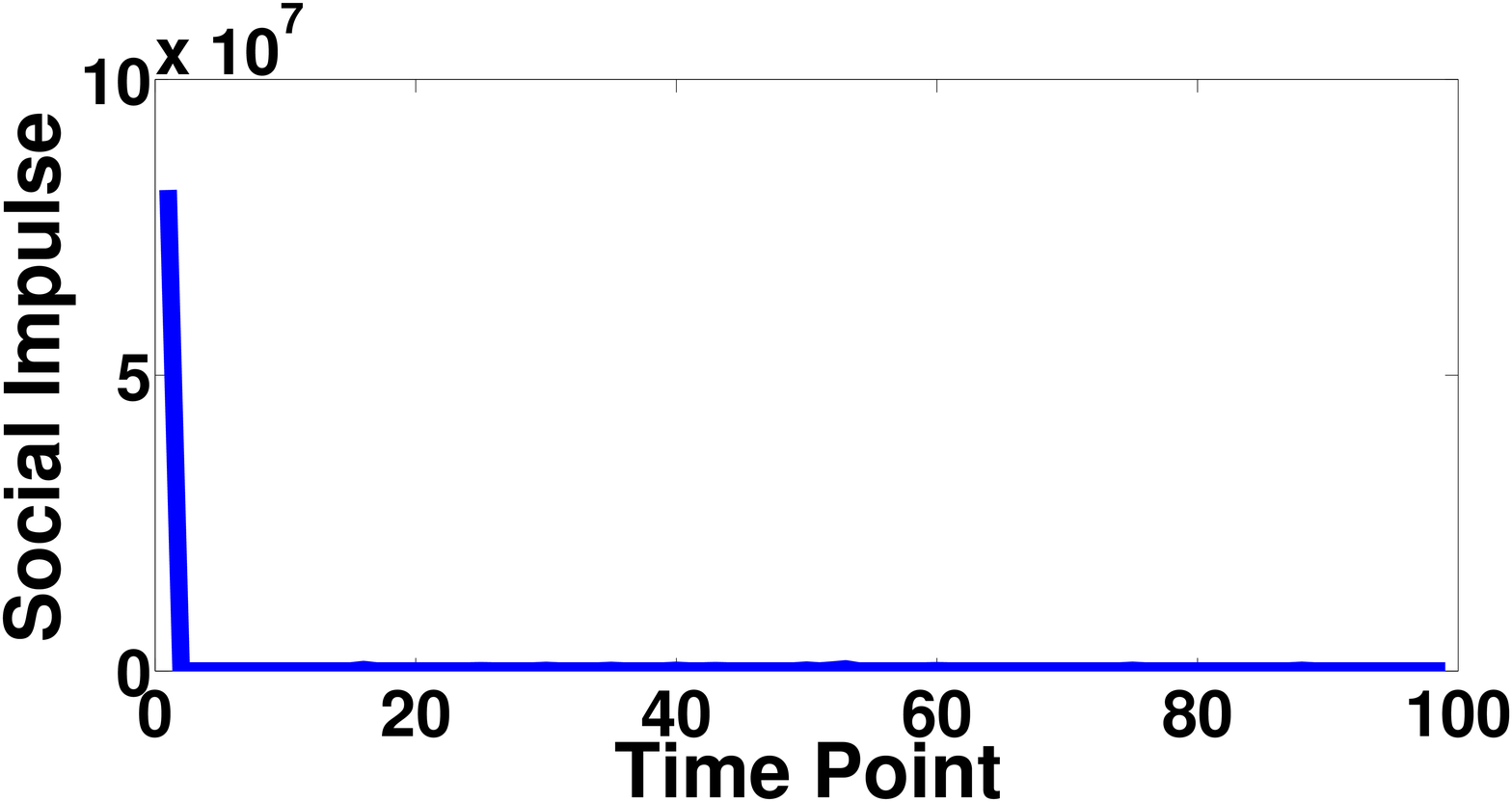}
		\label{fig: non_longevous_impuse}
	}
	\subfigure[Accumulated Views of a Longevous video] {
		\includegraphics[width=40mm, height=20mm]{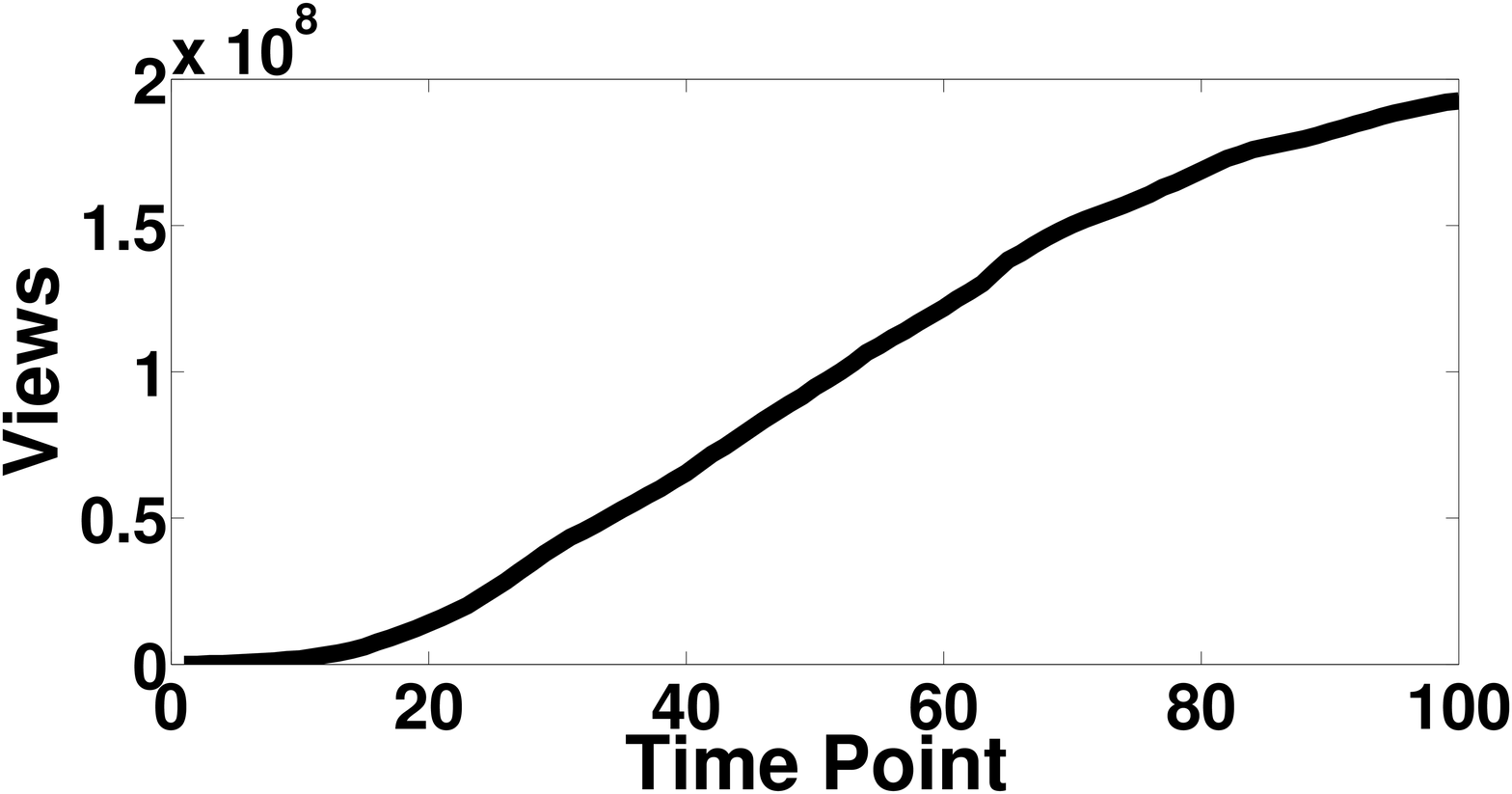}
		\label{fig: longevous}
	}
	\subfigure[Social Impulse History of a Longevous video] {
		\includegraphics[width=40mm, height=20mm]{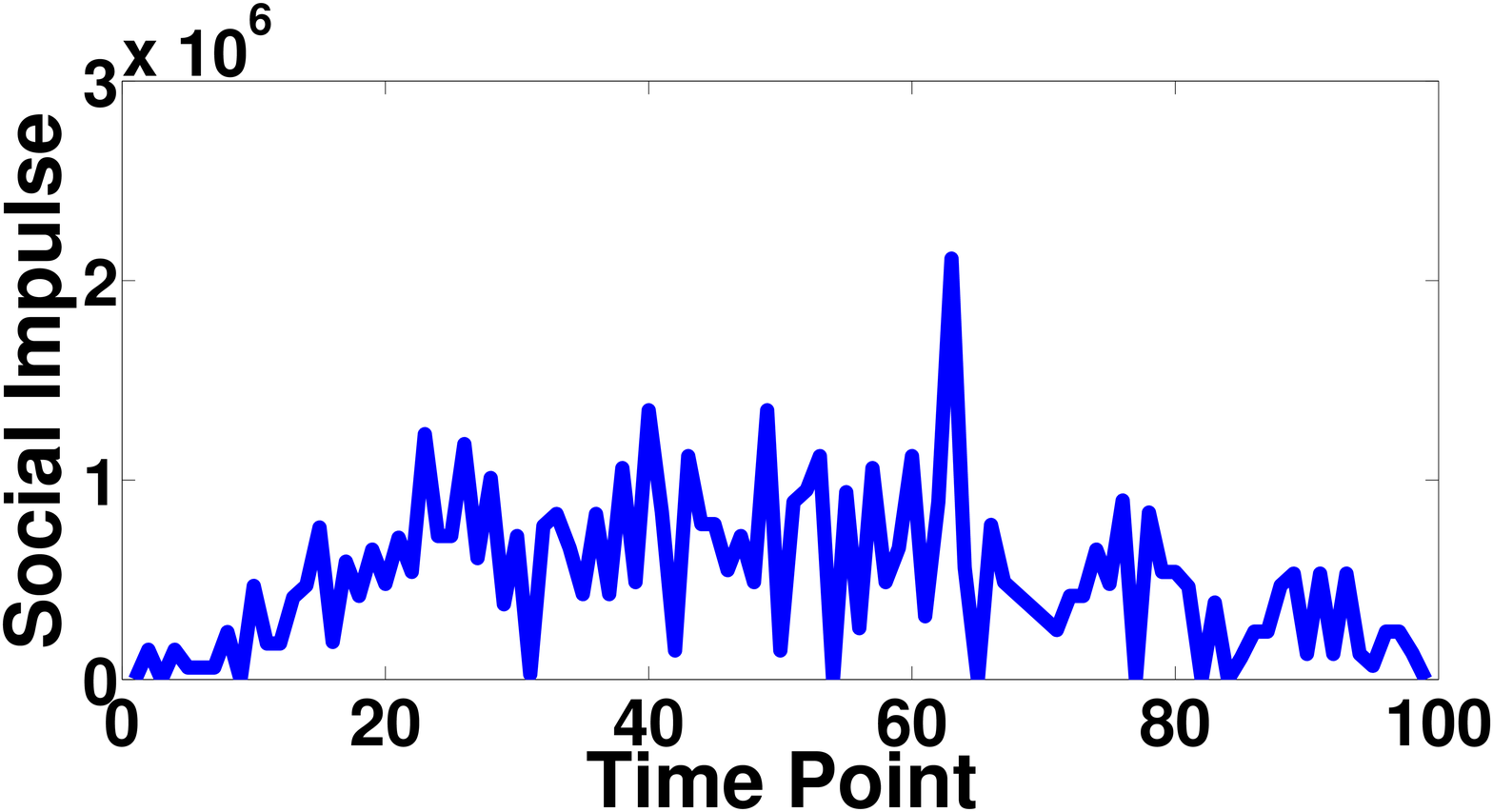}
		\label{fig: longevous_impulse}
	}
	
	\caption{Case Study of Longevous and non-Longevous Videos ($\gamma = 0.3$)}
	\label{fig: longevity_case_study}
\end{figure}

In addition, we provide a case study containing two picked videos whose scores are small (=13) and large (=90), respectively, when $\gamma=0.3$. The first video is in the Sports category and contains the playback of a goal in a soccer game. The second video is in the Music category and contains a live show of Bruno Mars, a famous American singer. These two videos are specifically selected as a case study because their published dates were close and both have received around 1.9 million views at the time when we collect data. However, as we can see in Figure \ref{fig: longevity_case_study}, the view count increase of the first video ceased at the very early stage and its history of social impulse contains only a very few spikes. The second video's view count, on the other hand, has a more steady growth pattern of view count and its history of social impulses is more active. The proposed longevity quantification method well captures the difference between these two growth patterns: it outputs a relatively small score (13) for the first video and a larger score (90) for the second one. In fact, similar to the videos in Figure \ref{fig: video_cases}, when we revisit these two studied videos in the September, 2013, we discover that the first video's total views are still around 1.9 million, yet the second video has received around 3 million total views. This again confirms that $L_1$ can effectively measure a video's long-lasting value.

\begin{figure}
	\centering
	\includegraphics[width=\linewidth]{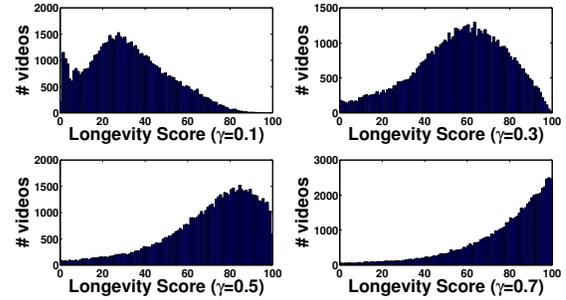}
	\caption{Longevity Score Distribution for Different $\gamma$}
	\label{fig:score_distribution}
\end{figure}

Next, we demonstrate the longevity score distribution of all videos in the crawled dataset, and how it is affected by the change of the value of $\gamma$. Figure  \ref{fig:score_distribution} shows the distribution under different assignments of $\gamma$. As we can see, the score distribution is roughly a Gaussian distribution, while $\gamma$ has a significant effect on the mean value of the calculated longevity scores. Such phenomenon can also be explained by our previous statements about the effect of changing $\gamma$ value: when $\gamma$ is smaller, we assume the system's response to a single social impulse decays more slowly, and the calculated social impulse vectors tend to be sparser. Since the longevous score is related to the number of social impulses whose strength are greater than a constant threshold ($\epsilon$), the mean of the longevity scores naturally decreases along with the decrease of $\gamma$. 

\begin{figure}[h]
	\centering
	\includegraphics[width=\linewidth]{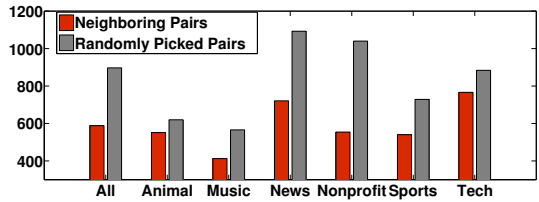}
	\caption{Average Squared Difference Between Videos}
	\label{fig: neighbor_scores}
\end{figure}

At last, as we introduced in the data collection section, we have organized a video network consisting of videos as nodes and "related video" relationship as edges. Based on this network, we further examine the difference between the scores of neighboring videos: for each edge, we calculate the squared difference of longevity scores between the two neighbors. Moreover, to complete the comparison, we also randomly draw some pairs of videos to calculate the squared difference, and the number of these pairs is exactly equal to the number of edges in the network. Figure  \ref{fig: neighbor_scores} shows the average squared difference of longevity scores between the neighboring paris and randomly drawn pairs on the entire video network, as well as the results of this test on the reduced subnetworks by using videos in the same category. The results show that the average squared difference between neighboring videos in the network is much smaller than the randomly picked ones, especially for some categories such as News and Nonprofit. We can easily understand such phenomenon by considering the nature of the video network: two connected videos have similar content and therefore have similar long-lasting value. The statistical findings in Figure \ref{fig: neighbor_scores} suggest us that video edges in the network are useful to infer unknown longevity scores from known ones. Such conclusion is crucial to solve the problem of predicting longevity scores.

\section{Prediction of Longevity Scores}

\subsection{Task Description}

In real life, the view count's history of a video may not be visible for everyone. As a result, the longevity cannot be directly quantified. This affects more seriously to third-party recommender systems, since they have almost zero information of a viewer's profile and only public information of videos. In fact, during the period of collecting videos, we observe that around 65\% of checked videos do not contain any view count's history. For certain categories, such as music, this number is even larger. The main reason behind this phenomenon is that many publishers of videos have turned off the function of providing public view count's history. Especially when the publisher of music videos are commercial companies, most of them choose not to unveil the history of view counts for their own interest. However, there are two types of data that are always available for every video: public features listed in Table \ref{tab: crawled_data_1} and the connection between videos through "related video" relationship. 

To sum up, in a real-life scenario, we are able to observe some videos' longevity scores (since their view count's histories are available), while others are unobservable and need to be predicted. Moreover, we are also able to obtain every video's public features and a network connecting different videos. Therefore, we can formulate the problem of predicting the longevity scores as a semi-supervised learning problem: a network mixed with labeled (known longevity scores) and unlabeled nodes (longevity scores need to be inferred). Afterwards, some interesting questions arise: (1) are we able to predict a video's longevity score by only using the public features? (2) How can we utilize the network of videos to make improvement? 

In the following paragraphs, we use the longevity score calculated when $\gamma = 0.3$ as the ground truth. The reason for doing this is that in Figure  \ref{fig:score_distribution}, we are able to observe a complete Gaussian-like distribution of the calculated longevity scores under $\gamma = 0.3$ and thus can make a clearer distinguish between longevous and non-longevous videos. In the following paragraphs, we will first introduce baseline models, as well as the proposed \emph{Gaussian Markov Random Field (GMRF)} model with \emph{Loopy Belief Propagation (LBP)} to solve this semi-supervised learning problem. Afterwards, we will use experiments on the crawled Youtube dataset to evaluate the proposed approach comparing to the two baselines.

\subsection{Baseline Models}

{\bf Regression Baseline\footnote{We tried classification algorithms by viewing scores as different labels. However, the result is generally much worse. For example, the mean squared error of a decision tree algorithm is more than 50\% worse than the linear regression method.}} A straightforward idea to predict longevity scores is to use regression on public features and assign videos to the closest possible scores according to the regression results. Therefore, except for "id" and "category" in Table \ref{tab: crawled_data_1}, we include all other numerical features into the regression task. To begin with, we first examine the Pearson correlation between each feature and the longevity score (Table \ref{tab:correlation}). As one can see, all features have positive correlation with the longevity score yet the correlation is not much significant, especially for the length of videos and "dislike" count. Such statistical finding informs us relying on a single feature, such as accumulated views, is not able to provide an accurate prediction of a video's longevity score. Therefore, we apply linear regression on all these features to predict more accurate scores. Later experiment will demonstrate the performance of this model.

\begin{table}
	\centering
	\begin{tabular}{c | c}
		Feature & Pearson Correlation Coefficient\\
		\hline
		view count & 0.1038\\
		\hline
		favorite count & 0.1116\\
		\hline
		average rating & 0.1034\\
		\hline
		length & 0.0618\\
		\hline
		"like" count & 0.1065\\
		\hline
		"dislike" count & 0.0464\\
	\end{tabular}
	\caption{Correlation Between Feature and Longevity Score}
	\label{tab:correlation}
\end{table}

{\bf K-Nearest Neighbor Baseline. } Previously, we have used statistical method to demonstrate the difference between neighbor videos are much smaller than random picked video pairs. Obviously, a natural method to predict a video's longevity score is to find the $k$ nearest videos whose longevity scores are observable and use the mode of these scores as the prediction result. In order to select the best $k$, we search in a range of values for each time we use the algorithm, and only report the best prediction result in later experiments. This K-Nearest Neighbor style approach and its extensions are constantly used in semi-supervised learning tasks~\cite{zhu2003semi, Ozaki2011}.

\subsection{Gaussian Markov Random Field (GMRF) Model}

Motivated by statistical analysis in the section 3, we believe {\bf Gaussian Markov Random Field (GMRF)} model is a wise choice for this problem. Therefore, we adopt the framework in~\cite{zhu2003semi} with necessary adjustments and extensions to fit our problem setting. First, for a node (video), say $i$, in the network, we represent it as an \emph{m}-dimensional vector ${\bf v_i} = (v_{i1}, v_{i2} ... v_{im})$, which stores the value of each feature and $m$ is the total number of features. In our case, the node's features are the same ones used in the linear regression baseline, which means \emph{m} = 6. For every edge connecting two nodes, say ${\bf v_i}$ and ${\bf v_j}$, we assign the following edge weight:
	\begin{equation}
		w_{i,j} = \exp \big( - \sum_{k=1}^{m} \frac{(v_{ik} - v_{jk})^2}{ \varsigma_k^2}\big)
		\label{eq: edge_weight}
	\end{equation}
where $\varsigma_k^2$ is the variance of the $k^{th}$ feature. By dividing $\varsigma_k^2$ in Equation (\ref{eq: edge_weight}), we are able to scale the effect of different features to the similar magnitude, since the numerical difference for some features, such as view count, are much larger than the others, such as video's average rating. Generally speaking, if two adjacent nodes are more similar to each other with respect to their features, Equation (\ref{eq: edge_weight}) tends to assign a greater weight to the edge connecting them. Of course, for any non-connected pairs of nodes, $w_{i,j} = 0$.

In order to simplify the model, we assume that each node's longevity score is drawn from the same distribution. Therefore, if  we denote a node's longevity score as $s_i$ ($s_i = \{0...100\}$), we define the following Gaussian form of node potential function, which is identical to every node:
	\begin{equation*}
		\psi_i(s_i) = \exp \big(-\frac{1}{2} \cdot \frac{(s_i - \mu)^2}{\sigma^2}\big)
	\end{equation*}
where $\mu$ and $\sigma^2$ are constants that can be estimated by fitting the distribution of longevity scores to Gaussian distribution. Intuitively, if two nodes connected by an edge with larger weight, we believe their labels (longevity scores) should be similar. As a result, the edge potential functions have the following quadratic form:
	\begin{equation*}
		\psi_{ij}(s_i,s_j) = \exp \big( -\frac{1}{2} \cdot  w_{i,j} (s_i - s _j)^2 \big) 
	\end{equation*}
where $s_i$ and $s_j$ are the longevity scores of the two connected nodes. It is easy to see that $\psi_{i,j}(s_i, s_j) = \psi_{j,i} (s_j, s_i)$, since all the edges in the network are undirected. Finally, the proposed pairwise GMRF is: 
	\begin{equation}
		p({\bf s} | \mu, \sigma) = \frac{1}{Z} \prod_{i}\psi_i(s_i) \prod_{i,j}\psi_{ij}(s_i, s_j)
		\label{eq: GMRF}
	\end{equation}
where $Z$ is the partition function. As we can see in Equation (\ref{eq: GMRF}), the node potential functions serve like prior knowledge about a node's label, while the edge potential functions depict the relationship between two adjacent nodes. Since $\mu$ and $\sigma$ are estimated constants and there is no additional parameter in the GMRF, we only need to focus on the inference of each unknown video's longevity score. 

\subsection{Inference by Loopy Belief Propagation (LBP)}

Unlike the baseline model proposed in~\cite{zhu2003semi}, we are not able to directly use the close form solution in matrix representation to performance inference on the proposed GMRF. The reason is that similar to social networks, video networks are also usually very sparse. As a consequence, the video network contains isolated nodes with no edge at all, as well as the connected components that only have unlabeled nodes. Therefore, these nodes will be singular points in the matrix form solution proposed in~\cite{zhu2003semi}. To overcome this problem, we adopt {\bf Loopy Belief Propagation (LBP)} on the GMRF to infer each unlabeled node's longevity score. 

For each node in every iteration, the LBP algorithm has two steps: sending message to its neighbors and updating its own beliefs. In the first step, if the score of a node, say $i$, is unknown, the message sent from it to an adjacent node, $j$, denoted as $m_{i \rightarrow j}$ is a vector calculated in the following way:
	\begin{equation}
		m_{i \rightarrow j}(s_j) = \sum_{s_i} \Big( \psi_i (s_i) \psi_{i,j} (s_i, s_j) \prod_{k \in nbr_i \backslash j} m_{k \rightarrow i} (s_i) \Big) 
		\label{eq: unknown_message}
	\end{equation}
where $nbr_i$ is the set of neighbors of node $i$, and $nbr_i \textbackslash j$ denotes $nbr_i$ discarding $j$. However, if the score of node $i$ is known, the message sent from it to an adjacent  node $j$ is calculated as follows:
	\begin{equation}
		m_{i \rightarrow j}(s_j) = \psi_i (s_i) \psi_{i,j} (s_i, s_j) \prod_{k \in nbr_i \backslash j} m_{k \rightarrow i} (s_i)
		\label{eq: known_message}
	\end{equation}
As shown in Equation (\ref{eq: unknown_message}) and (\ref{eq: known_message}), the message sent from one node to one of its neighbors is constructed by discarding the message received from the neighbor. The difference between Equation (\ref{eq: unknown_message}) and (\ref{eq: known_message}) is that when a node is unlabeled, the message sent from it is constructed based on the summation of potential functions on its all possible scores, while the message sent from a labeled node is generated by fixing its score to the observed true value. This process makes a labeled node, say $i$, like a radio station filtering out any incorrect information (i.e. ignoring $m_{j \rightarrow i}(s_i)$ when $s_i$ is not the true score) in the received message and continuously broadcasting the same type of message directly based on the truth.

In the second step of LBP, for each unknown node $i$, we update its beliefs (represented as a vector, $bel_i$) as follows:
	\begin{equation}
		bel_i(s_i) = \frac{1}{Z_i} \cdot \psi_i(s_i)\prod_{j \in nbr_i} m_{j \rightarrow i}(s_i)
		\label{eq: unknown_belief}
	\end{equation}	
where $Z_i$ is a normalization factor to ensure the summation of $bel_i(s_i)$ on $s_i=\{0...100\}$ exactly equals to 1. Of course, if a node's score is known, we directly fix its beliefs:	
	\begin{equation}
		bel_i(s_i) = \left \{
		\begin{array}{ll}
			1 & (s_i \text{ is equal to the true value})\\
			0 & \text{otherwise}
		\end{array}
		\right.
		\label{eq: known_belief}
	\end{equation}	
In each iteration of the LBP, we continuously apply Equation (\ref{eq: unknown_message}) $\sim$ (\ref{eq: known_belief}) on every node. The LBP stops until the beliefs of nodes stabilize or the number of iterations reaches a predefined bound to control the algorithm's running time. In our experiments, we set the bound to be 100, since the diameter of the network is only 27 (Table \ref{tab: crawled_data_2}). The final predicted score for an unlabeled node $i$ is the score having the maximum belief, which is argmax$_{s_i} \big(bel_i(s_i)\big)$. Algorithm \ref{alg: LBP} summarizes the proposed GMRF model with LBP.

\begin{algorithm}
	\caption{Loopy Belief Propagation on the proposed GMRF}
	\label{alg: LBP}
	\begin{algorithmic}[1]
		\REQUIRE ~\\
				 $G = (V, E, W)$: A video network. $V$ is the node set and $V = U \cup L$, where $U$ is the set of unlabeled nodes and $L$ is the set of labeled nodes ($U \cap V = \varnothing$). $E$ is the set of edges. $W$ stores the weight of each edge.\\
				 $T$: Iteration bound for LBP to control running time.
				 
		\ENSURE {$S$: \{$\hat{s_i}$ | $\hat{s_i}$ is the predicted score of node {\bf $v_i$}, $\forall$ {\bf $v_i$} $\in U$ \}}
				
		\STATE {Initialize all messages $m_{i \rightarrow j} (s_j) = 1$}
		\STATE {$t$=0}
		
		\WHILE {$bel_i$ is not stabilized and $t \leq T$}
			\FOR {each node {\bf $v_i$} $\in V$}
				\IF {{\bf $v_i$} $\in U$}
					\STATE{use (\ref{eq: unknown_message}) construct messages and (\ref{eq: unknown_belief}) to update $bel_i$.}
					\ELSE \STATE{use (\ref{eq: known_message}) construct message and (\ref{eq: known_belief}) to update $bel_i$.}
				\ENDIF
			\ENDFOR
			\STATE{t++}
		\ENDWHILE		
		
		\STATE{$S=\varnothing$}
		\FOR{each node {\bf $v_i$} $\in U$}
			\STATE{$\hat{s_i}$ = argmax$_{s_i} bel_i(s_i)$}
			\STATE{$S = S \cup \{\hat{s_i}\}$}
		\ENDFOR
		\STATE{{\bf return} $S$}
	\end{algorithmic}
\end{algorithm}

\subsection{Experiments}

Motivated by different real-world scenarios, we propose two experiments on the crawled Youtube dataset to evaluate the proposed GMRF and two baseline models.

{\bf Prediction on the whole network:} this task corresponds to the situation that we do not specify a particular video category. In this experiment, we randomly divide the nodes in the network into two sets. The nodes in the first set will be labeled, which means the longevity scores are known, and the nodes in the second set will be unlabeled. The tested ratios of the unlabeled nodes to all nodes are 0.1, 0.2, 0.3, 0.4 and 0.5. On the generated networks, we apply different algorithms to infer the longevity scores of all unlabeled nodes and record the Mean Squared Error (MSE) of the predicted scores for each algorithm. In order to be accurate, for each ratio of unlabeled nodes in network, we repeatedly to generate 50 networks, and the reported MSE is the average result on the 50 networks. 

{\bf Prediction on the categorized sub-networks:} Sometimes, we may have a target category of videos in mind. For example, a search engine should definitely recommend Sports rather than Animals video if the viewer's key words are all related to sports.
In this experiment, we explore whether it will be easier to predict the longevity scores in some categories than the others. We reduce the network to 6 sub-networks only containing nodes that belong to the same category, and the edge set is also reduced accordingly. Afterwards, similar to the prediction on the whole network, based on every sub-network, we randomly generate 50 networks with different unlabeled node ratios (0.1, 0.2, 0.3, 0.4, 0.5). Of course, since the prior knowledge of the longevity scores in different video categories may vary, when we apply the GMRF model, we need to change the $\mu$ and $\sigma$ in the node potential function accordingly. Similar to obtaining $\mu$ and $\sigma$ in the whole network, the updated $\mu$ and $\sigma$ can be computed by Gaussian fitting on longevity scores within the specific category. The reported MSEs are also the average values. 

\begin{figure}[th]
	\centering
	\includegraphics[width=0.8\linewidth, height=1in]{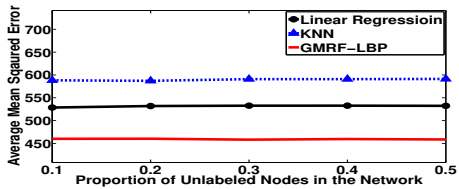}
	\caption{Average MSEs of Predicted Scores}
	\label{fig: prediction_result}
\end{figure}

For the first prediction task on all videos, average MSEs on networks having different proportions of unlabeled nodes are reported in Figure \ref{fig: prediction_result}. As we can see, the proposed GMRF with LBP significantly outperforms the other baselines. Moreover, the performance of compared algorithms are all stable despite the change of the ratio of unlabeled nodes. For linear regression model, such phenomena is easy to explain: since labeled and unlabeled nodes are randomly divided, we expect the distribution of labeled nodes' features and longevity scores are all persistent despite the change of labeled ratio. As a result, the trained model may be very alike and thus have similar prediction power. Similar to the linear regression, we observe that reducing the proportion of labeled nodes is not going to affect the performance of network-based methods (KNN and GMRF) either. This demonstrates that such methods are also robust in the prediction task. In fact, for a specific semi-supervised learning task, such phenomenon is not unique~\cite{li2012mining}.

\begin{figure}[th]
	\centering
	\subfigure[Animals]{
		\includegraphics[width=40mm]{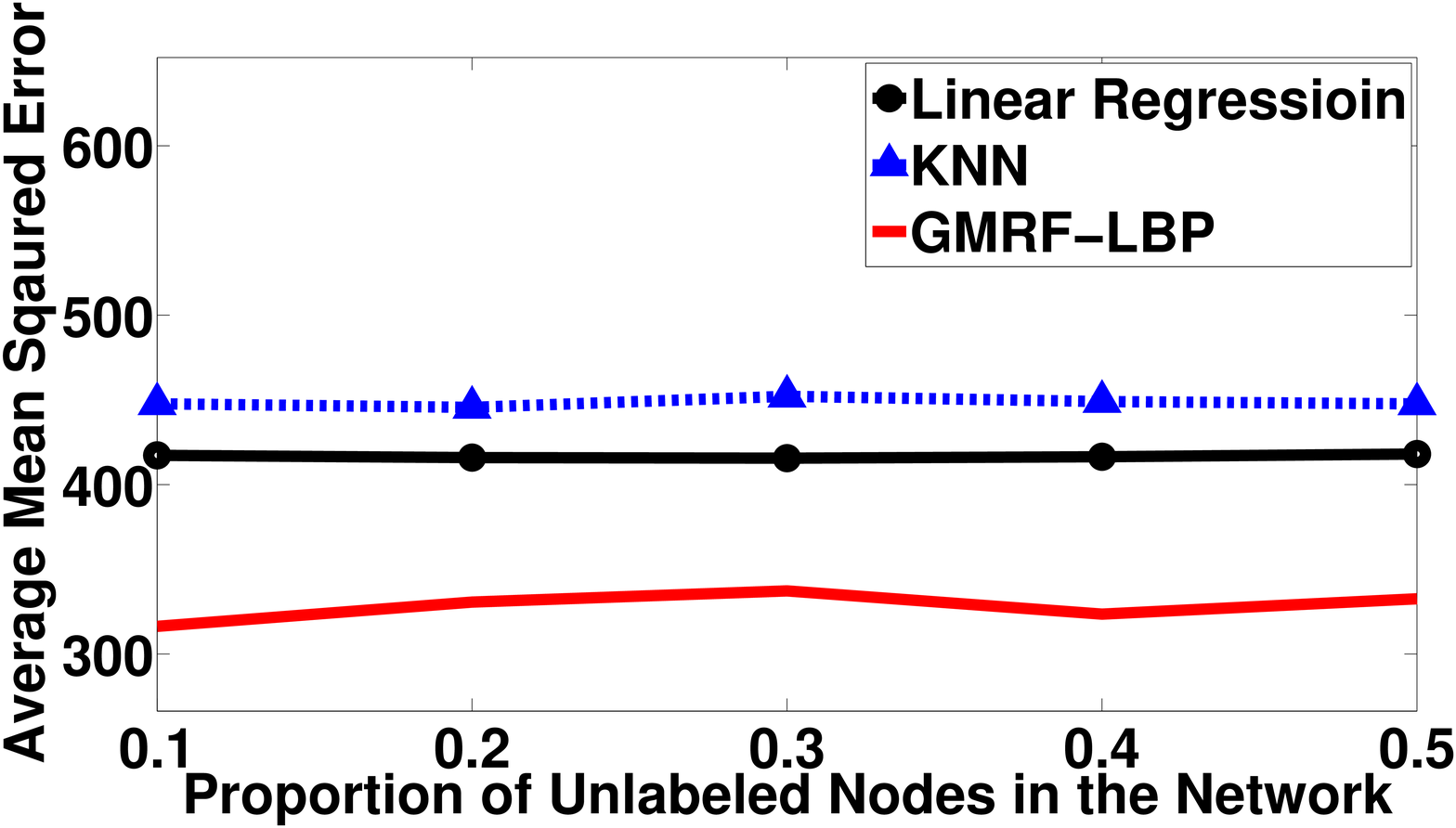}
	}
	\subfigure[Music]{
		\includegraphics[width=40mm]{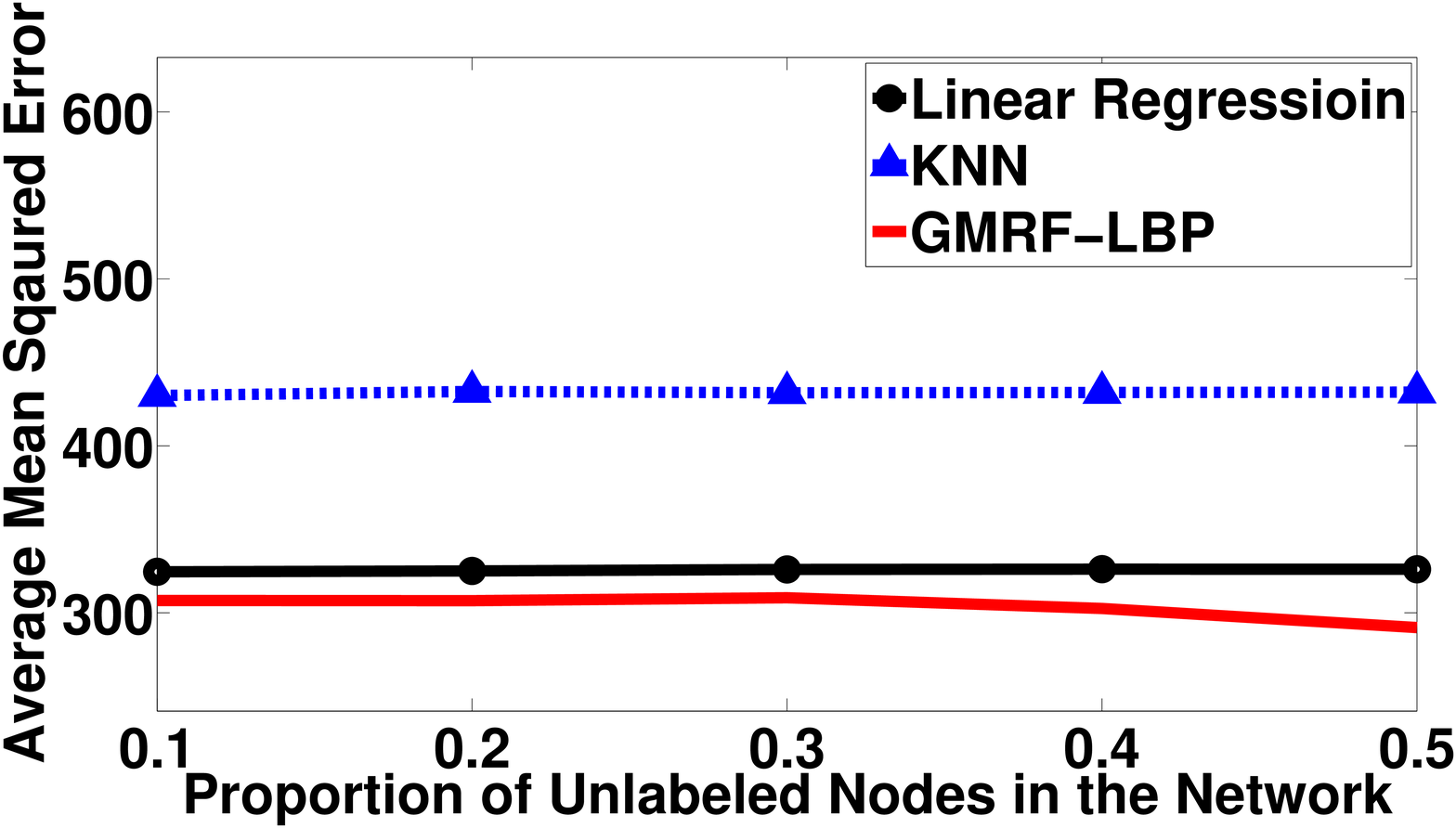}
	}
	\subfigure[News]{
		\includegraphics[width=40mm]{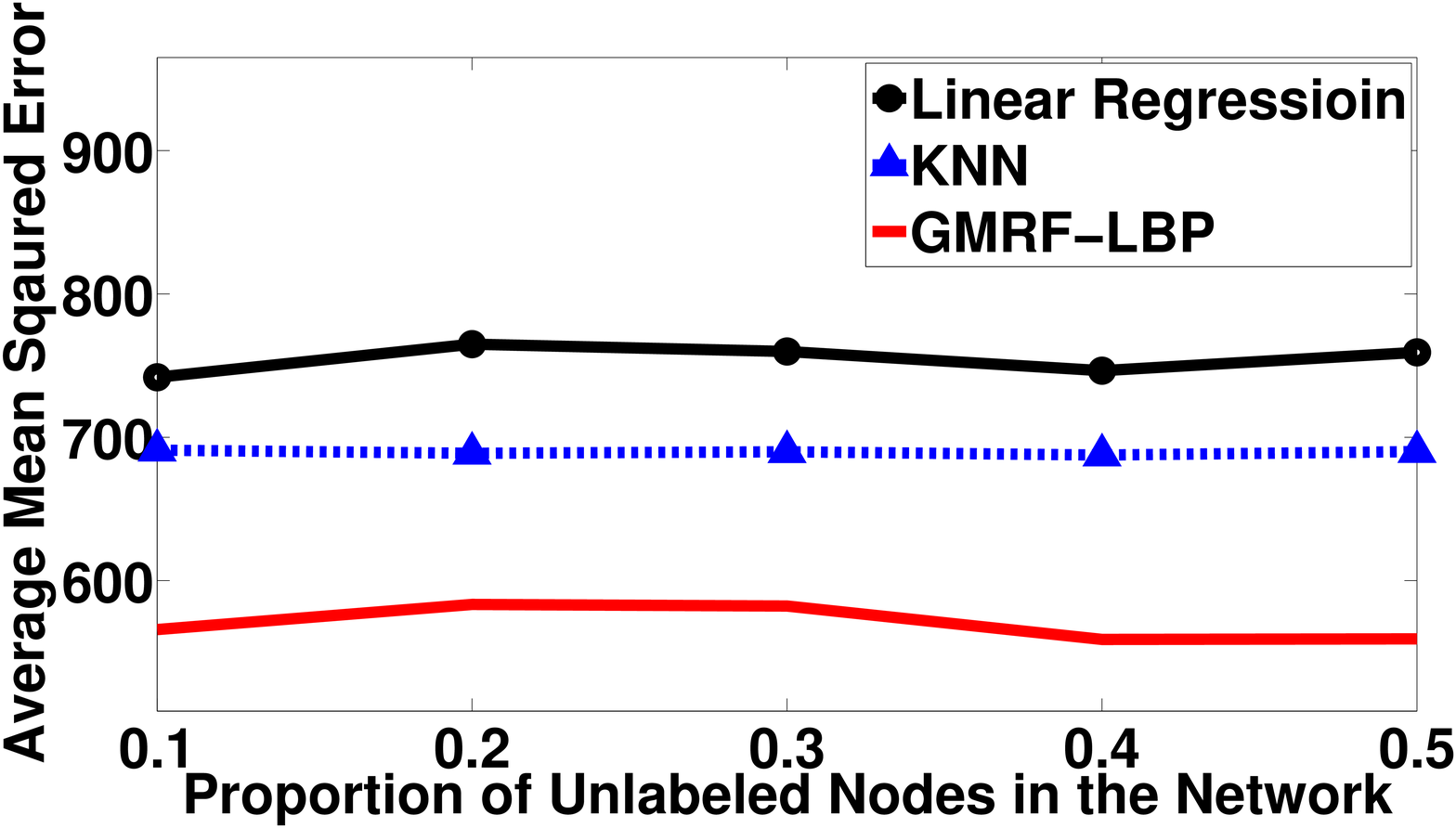}
	}
	\subfigure[Nonprofit]{
		\includegraphics[width=40mm]{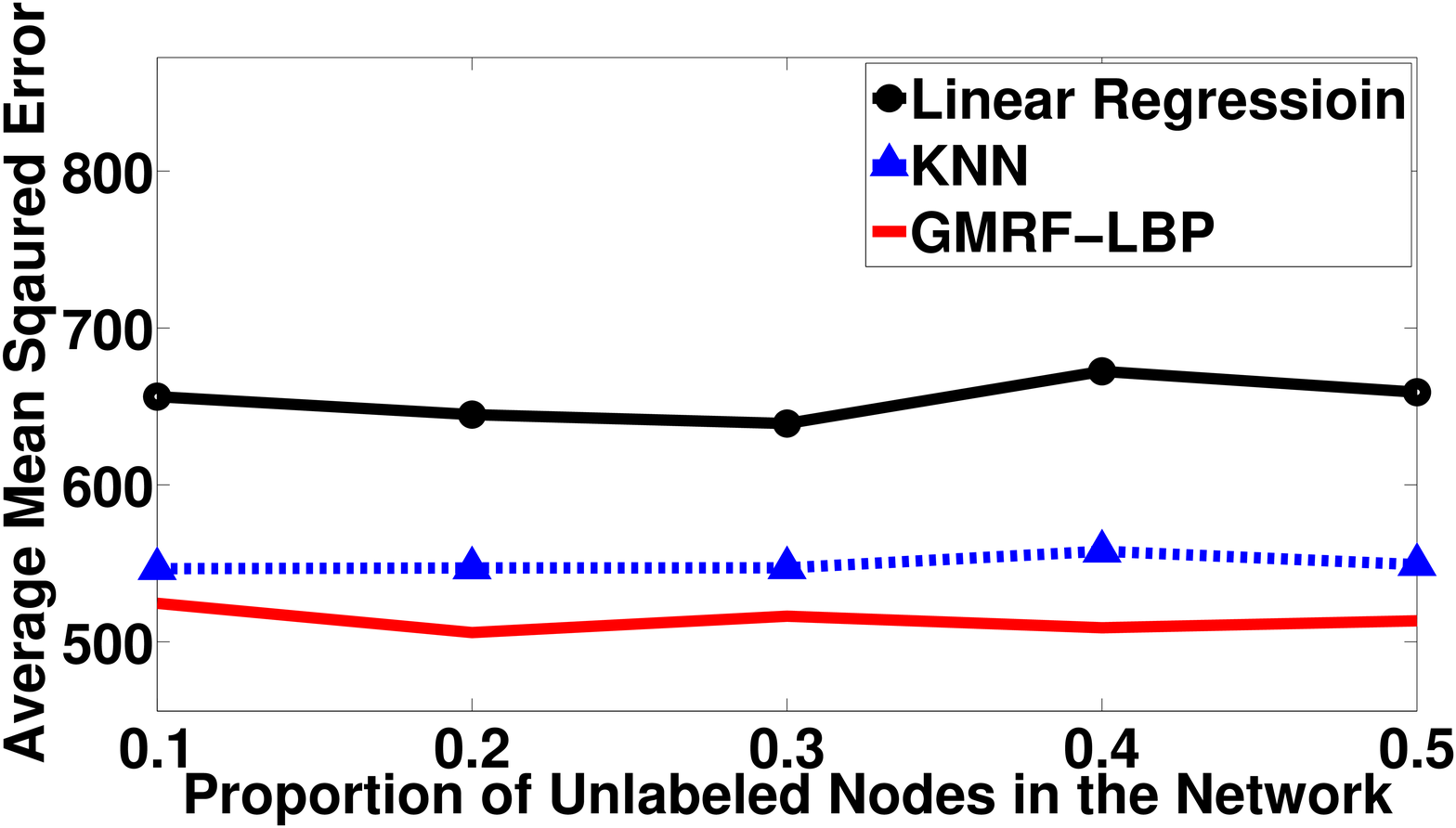}
	}
	\subfigure[Sports]{
		\includegraphics[width=40mm]{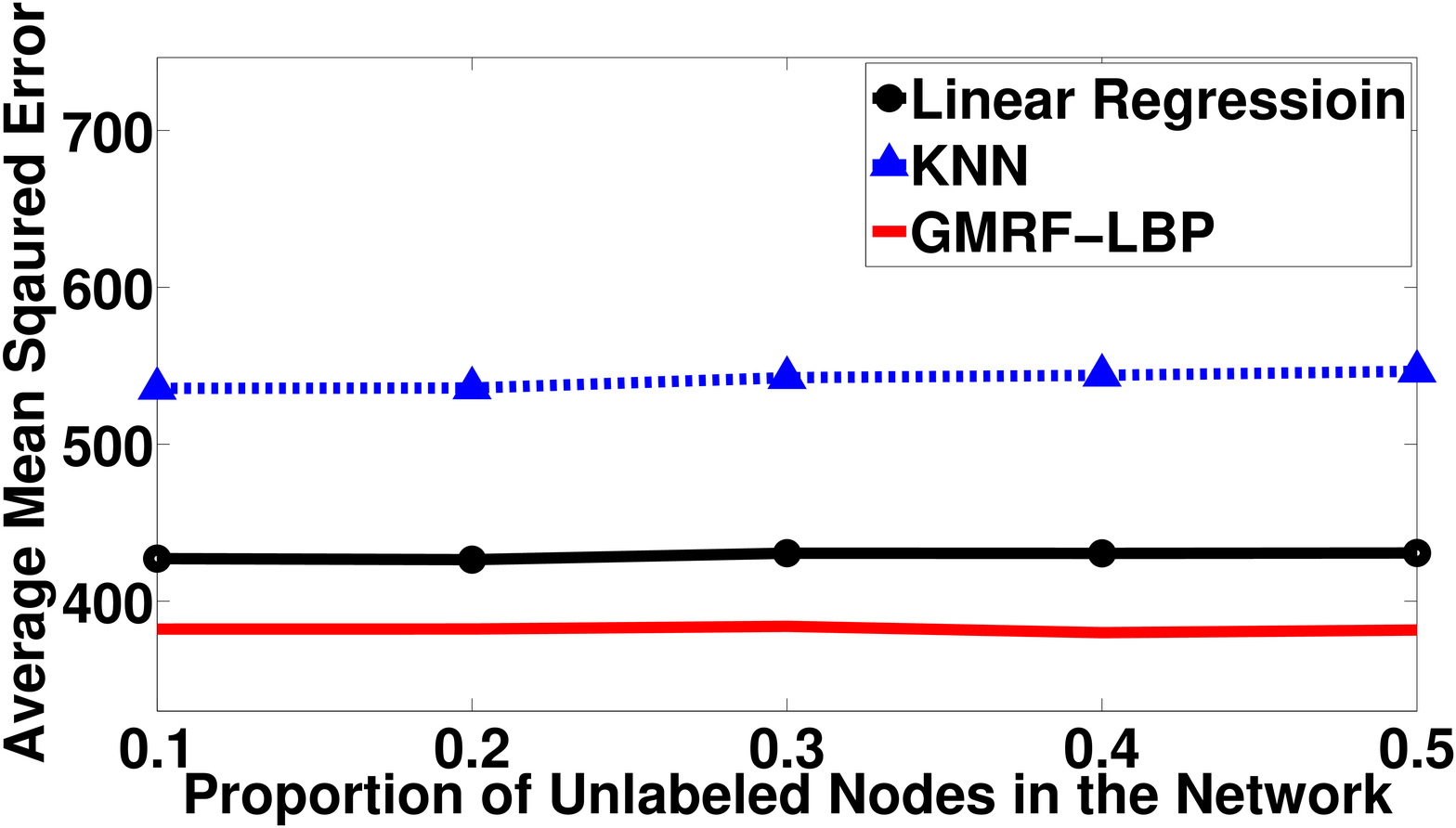}
	}
	\subfigure[Technology]{
		\includegraphics[width=40mm]{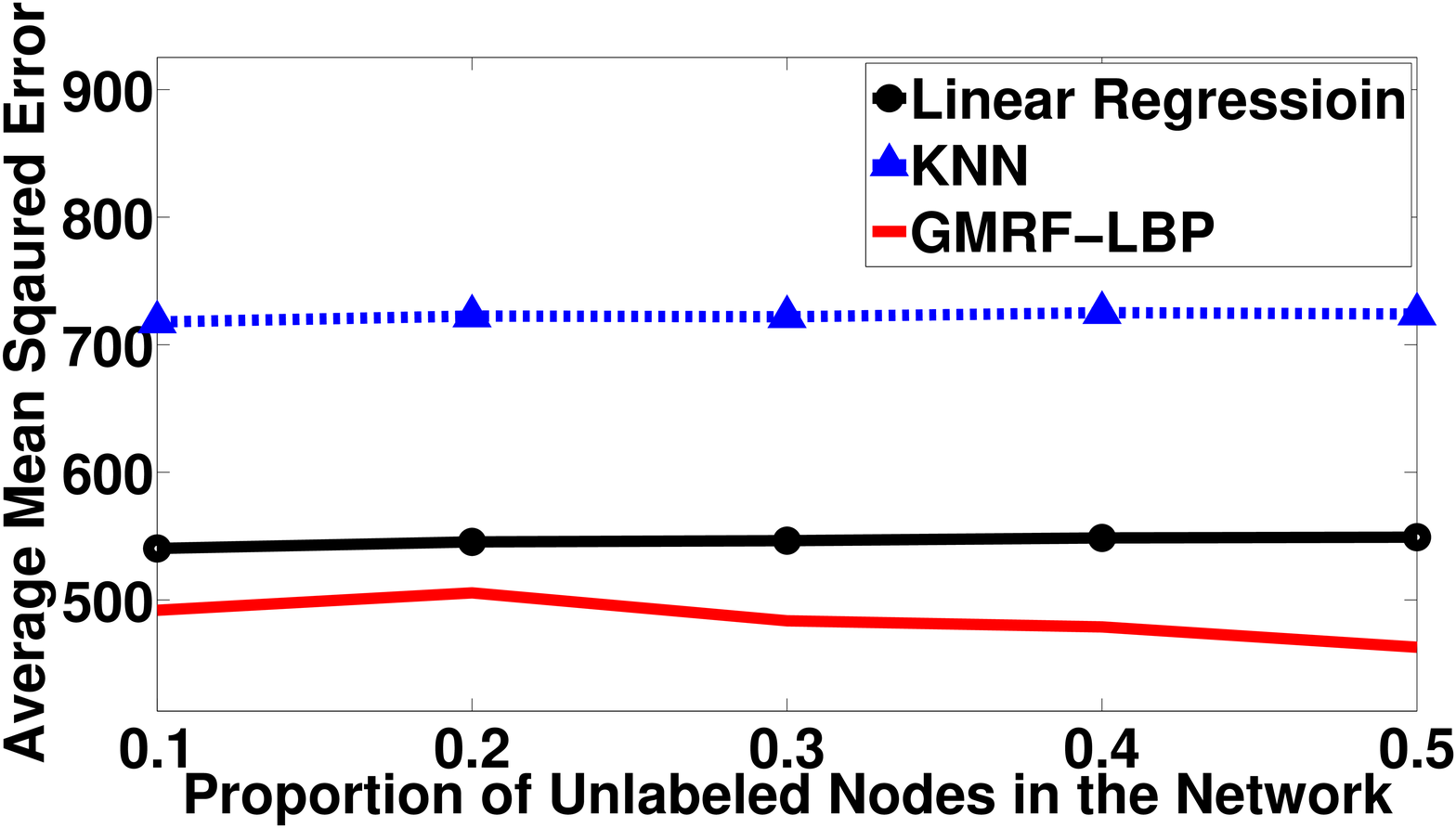}
	}
	
	\caption{Average MSEs of Predicted Scores on Categorized Sub-networks}
	\label{fig: cate_prediction_result}
\end{figure}

Similar to the whole network, the results on the categorized sub-networks, which are reported in Figure \ref{fig: cate_prediction_result}, also confirm that the GMRF algorithm is the best one among all three approaches. What is more, by comparing the results across different sub-networks, we discover that predicting longevity scores for videos in Animals, Music and Sports category have lower average mean squared errors than the others. Another interesting finding is that although the KNN algorithm is usually worse than the linear regression, it is better in some specific categories, such as News and Nonprofit.

\section{Related Work}

Topics related to the popularity of online videos start to attract the attention from researchers in recent years. Early articles conducted many statistical analysis on the online videos website such as Youtube~\cite{Youtube_statistics, cha2007tube, Youtube_workload, 4801529, 5277844}. Due to the fact that most of published videos in Youtube are entirely user generated content, the statistics, therefore, demonstrate novel phenomena on major indices that is distinguished from the videos in traditional medias. In previous research, "popularity" is usually a video's total views~\cite{cha2007tube, Szabo_2010, Pinto2013, Youtube_popularity_growth}, which means it evaluates a video's value at a certain time point. Motivated by the potential advertising profit, articles such as~\cite{cha2007tube, Szabo_2010, Pinto2013} attempt to use different methods and feature sets to predict a video's popularity, a.k.a number of views, in the future. Unlike popularity in this paper, longevity evaluates a video's long-term value based on its entire history rather than a single point. Moreover, a trend in the society usually causes a decaying effect on a social media system. A good example are the Tweets related to one topic trend in the Twitter network~\cite{asur2011trends} and the disperse of a news event through an online social network~\cite{kong2012ranking}. As a result, similar effect causes that a single latent trend may stimulate continuous yet decaying increase of a video's views. Due to such effect, we believe it is more reasonable to measure the longevity based on the history of latent trends, a.k.a social impulses, rather than the history of view count increase. 

To derive latent social impulses, we presume that a social impulse has an exponentially tailed effect on a video's views. In fact, similar exponential functions are often used to describe an event's effect to a online service user. For example, \cite{Saito2010} uses this type of function to model the delay time of a user's response to a social trend or event and \cite{Iwata2013} uses it to model the influence of a user's shared event to his/her friends. Similarly, such exponential influence is also widely exist in the area of signal processing~\cite{Hayat92, smith1998localization}. Analogously, we view an online video website as a digital signal filter with an exponential impulse response function. The system takes social impulses as input signals and outputs the view count's history. However, unlike predicting a system's outputs with inputs, which is very common in the signal processing area, we need to estimate the system's inputs with known outputs. By using least squared error estimation, deriving social impulses is formulated as a convex optimization problem.

In the part of predicting a video's longevity score, the video's network connected by "related videos" relationship is very crucial. Previously, articles like~\cite{Youtube_social_network, Youtube_peer_to_peer_sharing} studied such "social network" of Youtube videos. Unlike human spontaneously form the link to each other, videos are connected according to their content. Interestingly, researchers found that the video's network has many common points with human social networks, such as the "small world" phenomenon~\cite{Youtube_social_network}. With the help of the video's network, we are able to formulate the prediction problem of a video's longevity score as a semi-supervised learning task. Therefore, we are able to adopt the framework in~\cite{zhu2003semi}, which uses Gaussian Markov Random Field and harmonic functions. However, unlike the original model only has edge potential functions, we also add node potentials by using prior knowledge derived from statistical analysis. Moreover, the original model deals with a general prediction problem, while our problem has a specific setting: the network is very sparse. Because of this phenomena, the close form solution proposed in~\cite{zhu2003semi} will have many singular points which cannot be estimated. Therefore, we apply Loopy Belief Propagation algorithm~\cite{murphy1999loopy} on the proposed model to infer the unknown longevity scores. According to the experiments, such adjustments make the proposed method better fits our problem and yields promising results.

\section{Conclusion}

As a new thriving force in the online world, video sharing websites largely enrich online media and generate huge opportunities for business. In this paper, we have discussed the benefit of evaluating and scoring the long-lasting value, a.k.a. longevity, for both third party companies and video websites. A longevous video tends to keep attracting society's attention, and thus has higher view count increase in the future. Based on the crawled data from Youtube, we discuss the latent temporal bias in a video's view count's history, which is the decaying effect that any event/trend may impose on a social media system. Through examples of real online videos, we argue the necessity to correct such bias before we evaluate a video's longevity. Social impulses are derived latent factors stimulating the increase of a video's view count. The proposed longevity calculation is based on the history of social impulses, since it can reduce the influence of the latent temporal bias and therefore to be more reasonable. In order to derive social impulses behind a video's view count's history, we view the video website as a digital signal filter and formulate this task as a convex optimization problem. As a result, the history of social impulses can be easily derived and the longevity thus can be quantified. However, unfortunately, not every online video has a public view count's history. Therefore, it becomes a challenge for third parties to evaluate longevity of all videos with limited public information. In the second part of this paper, we exploit the possibility to predict a video's longevity when the view count's history is missing. We formulate this problem as a semi-supervised learning task and use public features of videos, as well as the network constructed by "related videos" relationship to solve it. The proposed method is a Gaussian Markov Random Field model with Loopy Belief Propagation algorithm. At last, the experiments on the crawled dataset show that the proposed model significantly outperforms other baseline methods.

\bibliographystyle{abbrv}
\bibliography{reference} 

\end{document}